\documentstyle[epsfig,12pt]{article}
\newlength{\dinwidth}
\newlength{\dinmargin}
\newcommand{\qsl}{q \hspace{-5pt} / }

\def\vub{V_{ub}}
\def\vcb{V_{cb}}
\def\vtb{V_{tb}}
\def\vuqst{V_{uq}^*}
\def\vcqst{V_{cq}^*}
\def\vtqst{V_{tq}^*}

\newcommand{\ba}{\begin{array}}
\newcommand{\ea}{\end{array}}
\newcommand{\be}{\begin{equation}}
\newcommand{\ee}{\end{equation}}
\newcommand{\bea}{\begin{eqnarray}}
\newcommand{\eea}{\end{eqnarray}}

\def\bra{\langle}
\def\ket{\rangle}

\def\a{\alpha}
\def\b{\beta}
\def\g{\gamma}
\def\d{\delta}
\def\e{\epsilon}
\def\p{\pi}

\def\l{\lambda}
\def\m{\mu}
\def\n{\nu}
\def\G{\Gamma}

\def\to{\rightarrow}
\begin{document}
\thispagestyle{empty}
\addtocounter{page}{-1}
\begin{flushright}
DESY 97-126\\
hep-ph/9707251\\
July 1997
\end{flushright}
\vspace*{1.8cm}
\begin{center}  
{\large\bf An analysis of two-body non-leptonic\\
 $B$ decays involving light mesons in the Standard Model}
\end{center}
\vspace*{1.0cm}
\centerline{\large\bf A. Ali and C. Greub}
\vspace*{0.5cm}   
\centerline{\large\bf Deutsches Elektronen Synchrotron DESY, Hamburg,
                                                      Germany}   
\vspace*{1.0cm}
\centerline{\Large\bf Abstract}
\vspace*{1cm}
  We report a theoretical analysis of the exclusive non-leptonic 
decays of the $B^\pm$ and $B^0$ mesons into two light mesons,
some of which have been measured recently by the CLEO
collaboration. Our analysis is carried out in the context of an effective
Hamiltonian based on the Standard Model (SM), using 
next-to-leading order perturbative QCD calculations. We 
explicitly take into account the $O(\a_s)$ penguin-loop diagrams
of all four-Fermi operators and the $O(\a_s)$ tree-level diagram
of the
chromomagnetic dipole operator, and give a prescription
for including their effects in non-leptonic two-body decays. 
Using a factorization ansatz for the hadronic matrix elements, 
we show that existing data, in particular the branching   
ratios ${\cal B} (B^\pm \to \eta^\prime K^\pm)$, ${\cal B} (B^\pm \to
\pi^\pm K^0)$, ${\cal B} (B^0 (\overline{B^0}) \to \pi^\mp K^\pm)$,
and ${\cal B} (B^\pm \to \omega h^\pm) ~(h^\pm=\pi^\pm, K^\pm)$, can be 
accounted for in this approach.  Thus, theoretical scenarios with a 
substantially enhanced Wilson
coefficient of the chromomagnetic dipole operator (as compared to the SM)
and/or those with a substantial color-singlet $c\bar{c}$ component in
the wave function of $\eta^\prime$ are not required by these data.
 We predict, among other decay rates, 
the branching ratios for the decays 
$B^0 (\overline{B^0}) \to \pi^\pm \pi^\mp$ 
and $B^\pm 
\to \pi^0 \pi^\pm$, which are close to the present experimental limits. 
Implications of some of these
measurements for the parameters of the CKM matrix are presented. 

\vspace*{1.5cm}
\centerline{(Submitted to Physical Review D)}

\newpage
\section{Introduction}
\label{sec:introd}
\setcounter{equation}{0}

 Recently, the CLEO collaboration has reported
first measurements in a number of exclusive decays, 
$B^\pm \to \eta^\prime K^\pm$, $B^\pm \to
\pi^\pm K^0$, $B^0(\overline{B^0}) \to \pi^\mp K^\pm$, 
$B^\pm \to \omega h^\pm
~(h^\pm=\pi^\pm, K^\pm)$, $B^\pm \to K^0 h^\pm, ~B^\pm \to \pi^0 h^\pm$,
$B^0(\overline{B^0}) \to h^\pm \pi^\mp$,
 and the inclusive decay
$B^\pm \to \eta^\prime +X$ \cite{Behrens97,Jsmith97,Wuerthwein97},
which involve the so-called QCD penguins. In addition, a number of related
decays such as $B^0(\overline{B^0}) \to \pi^\mp \pi^\pm$ and 
$B^\pm \to \pi^\pm \pi^0$ are on the verge of measurement 
\cite{Wuerthwein97}. On the theoretical side, considerable effort 
has gone into studies of non-leptonic weak decays in terms of 
estimating decay rates \cite{BSW87} - \cite{NS97} and
the inherent direct and indirect CP asymmetries 
\cite{KPS94},  \cite{BSS79} - \cite{Ciuchini97-2}.
Since the first measurements of the above-mentioned decays,
theoretical interest in this subject has surged and recent literature is 
rife with all kinds of interesting interpretations of data, both within and 
beyond the SM \cite{HZ97} - \cite{FM97-2}. Of these, the decay mode 
$B^\pm \to \eta^\prime K^\pm$ is conspicuous due to its reported high
branching ratio ${\cal B} ( B^\pm \to \eta^\prime K^\pm )= (7.1 
^{+2.5}_{-2.1} \pm 0.9) \times 10^{-5}$ \cite{Wuerthwein97}.

The standard theoretical framework to study non-leptonic $B$ decays is 
based on the effective Hamiltonian
approach, which allows to separate the short- and long-distance 
contributions in these decays using the Wilson operator product expansion 
\cite{Wilson}. QCD perturbation theory is then used in
deriving the renormalization group improved short-distance
contributions \cite{AM74}.
This program has now been carried out up to and including the
next-to-leading order terms \cite{CMM96,Burasetal92}, but
the long-distance part in the two-body hadronic decays $B \to M_1 M_2$
involves  the transition matrix elements $\langle M_1 M_2 |  
O_i |B \rangle$ at a typical hadronic scale, where $O_i$ 
is an operator in the effective Hamiltonian (see below). Calculating
these matrix elements from  first principle is a true challenge in theory
which remains to be met. In view of this,
a number of approximate schemes has been put forward. The one we use here 
is based on the idea of factorization
\cite{Feyn65,EGN75}, in which the final state interactions (FSI) are 
assumed absent, and hence the hadronic matrix elements in the decay $B 
\to M_1 M_2$  
factorize into a product of two comparatively more tractable matrix elements.
These are then taken either from data or calculated in well defined 
theoretical contexts, such as QCD sum rules and potential models
\cite{BSW87,NS97}, \cite{ISGW}-\cite{EFG97}.
This framework does remarkably well
in accounting for non-leptonic two-body $B$ decays involving the
current-current operators
$O_{1,2}^{c}$ \cite{BSW87,Chengsoares} (see section 2 for definition). 
Recent analyses have shown that data on two-body  non-leptonic $B$
decays on the so-called heavy-to-heavy transitions, such as
$B \to (D,D^*) h$, $B \to (D_s,D_s^*) D,~B \to J/\psi h$ (with $h$ being a
light hadron),  can be described in terms of
two phenomenological parameters, $a_1$ and $a_2$ \cite{BSW87}, whose values 
seem to be universal \cite{BHP97,NS97}. Techniques
based on HQET \cite{HQET} allow in some limited cases to ``derive" such
factorization properties \cite{Bj89} and yield results which are in
agreement with data.

Motivated by the phenomenological success of factorization
in the heavy-to-heavy non-leptonic $B$ decays,
we would like to pursue further this framework in the domain of the so-called
heavy-to-light transitions, $B \to h_1 h_2$, where $h_1$ and $h_2$ are
light hadrons. The
recently measured $B$ decays \cite{Behrens97,Jsmith97,Wuerthwein97} belong
to this category and they should be analyzed on their own, without
prejudice about the suggested values of the effective parameters from
the heavy-to-heavy transitions. The decays $B \to h_1 h_2$ in
most cases involve mixing among the current-current,  QCD-penguin 
and the chromomagnetic operators. Our hope is that once perturbative QCD
corrections are taken into account,
these decays may allow themselves to be described in terms of a few 
phenomenological parameters.
 Related work along these lines
concerning QCD penguins in non-leptonic $B$ decays has been done
prior to this analysis \cite{Deandrea93,KPS94}, which we make
use of here, improve upon and extend.

Our analysis is based on the following three main ingredients:
\begin{itemize}
\item 
We work at next-to-leading logarithmic (NLL) precision, taking into account
the $O(\a_s)$ one-loop penguin-type diagrams of all four-Fermi operators
in the effective Hamiltonian and some process-independent parts of the 
vertex correction diagrams associated
with these four-Fermi operators.
We also take into account the effect of the 
$O(\a_s)$ tree-level matrix element associated with the 
chromomagnetic dipole operator via the process $b \to sg \to s \bar{q}'q'$.

\item To calculate the hadronic matrix elements, we  propose a simple 
factorization ansatz which allows to include
the effects of the $O(\a_s)$ matrix elements just discussed above.

\item In calculating  $B$ decays involving an $\eta^\prime$ or 
$\eta$ meson, such as $B^\pm \to \eta^\prime K^\pm$ and $B^\pm \to \eta 
K^\pm$,
 we include the contribution from the decay $b \to s (c\bar{c}) 
\to s(\eta, \eta^\prime)$ \cite{Berkelman,Smith}.
 The required decay constants and mixing parameters
are estimated using data on the radiative decays
 $J/\psi \to \eta_c \gamma, \eta^\prime \gamma, \eta \gamma$ and the 
two-photon decays of the $\eta, \eta^\prime$, and $\eta_c$ \cite{PDG96}.
Concerning $(\eta,\eta')$-mixing, we discuss both the conventional (one 
mixing-angle) formalism \cite{GK87,VH97} and the one involving two 
mixing-angles in this sector, which is suggested by the $1/N_c$-improved
$U(3) \otimes U(3)$ chiral perturbation theory framework
\cite{Leutwyler97,HSLT97}. Since the latter formalism is also favoured
by a recent phenomenological analysis \cite{FK97} of the
data on the $\eta \gamma$ and 
$\eta' \gamma$ form factors \cite{CLEOetagamma}-\cite{CELLOetagamma}, we use 
it in our estimates for $B$ decays involving $\eta'$ and $\eta$-meson.
\end{itemize}

We would like to make a number of remarks pointing out the
overlaps and differences with earlier analyses and
explaining our factorization ansatz.

Concerning the QCD-perturbative part, we note that our calculations 
come close to the derivation given in \cite{KPS94} but are more complete
as far as the NLL contribution is concerned.
We find that the NLL improvements implemented by us
reduce the scale dependence in various non-leptonic decay rates. This result 
is in line with what has been demonstrated in the radiative decays $B \to X_s +
\gamma$ in the same accuracy \cite{GHW96}. Further, 
the complete NLL contribution is important numerically, both compared to 
the leading order result and the NLL result obtained by keeping only 
the charm penguin contributions from the operators $O_{1,2}^{c}$.
We show this quantitatively in the context of the branching
ratio ${\cal B} (B^\pm \to K \pi^\pm)$, comparing it with the estimates
of the same based on keeping only the $O_{1,2}^c$ penguins \cite{FM97-1}.

Concerning the second point noted above, we remark that  our factorization 
prescription  introduces just one free parameter, called
$\xi$, which is supposed to compensate for the neglect of color
octet-octet contribution in evaluating the hadronic matrix elements in
the heavy-to-light sector $B \to h_1 h_2$.
This modifies the strength of the effective coefficients 
$a_1,...,a_6$  from their perturbatively calculated values (see section 3).
Clearly, this is the simplest ansatz and may have to be modified
eventually as more precise data on heavy to light $B$ decays become 
available.

We discuss the last point mentioned above concerning the decays
 $B^\pm \to \eta^\prime (K^\pm, K^{\ast \pm})$ and $B^\pm \to \eta
(K^\pm,K^{\ast \pm})$.
Expressing the charm quark content in the $\eta^\prime$ meson in terms of the 
matrix element $\langle \eta^\prime 
| \bar{c} \gamma_\mu \g_5 c | 0 \rangle = - i 
f_{\eta^\prime}^{(c)} q_\mu$, we find using data on the $J/\psi \to \eta_c 
\gamma$ and $J/\psi \to \eta^\prime \gamma$ decays that
 $|f_{\eta^\prime}^{(c)}| \simeq 5.8$ MeV. 
 The corresponding decay constant for $\eta$ meson is estimated
to be $|f_\eta^{(c)}| \simeq 2.3$ MeV in the conventional 
$(\eta,\eta')$-mixing formalism and $|f_\eta^{(c)}| \simeq 0.93$ MeV in
the $1/N_c$-improved approach.
The decrease in the value of $|f_\eta^{(c)}|$ reflects the small value of
the singlet mixing angle $\theta_0$, which makes the $\eta$ an almost pure
octet state \cite{Leutwyler97}, hence also reducing the
$c\bar{c}$ component of the $\eta$-meson. 
Our estimate $|f_{\eta^\prime}^{(c)}| \simeq 5.8$ MeV is to be contrasted 
with the range $f_{\eta^\prime}^{(c)} = 
(50 - 180)$ MeV obtained in \cite{HZ97}.  (Likewise, we find
$|f_{\eta^\prime}^{(c)}/f_{\eta^\prime} | \simeq 0.08$, which is also
an order of magnitude smaller than the one given in \cite{SZ97}.)
We note that our estimate of $|f_\eta^{(c)}|$ is consistent with the
bounds $-65 ~\mbox{MeV} \leq f_\eta^{(c)} \leq 15 ~\mbox{MeV}$, which
have been obtained in the meanwhile from an analysis of the $Q^2$-dependence 
of the
electromagnetic form factor of  $\eta'$ \cite{FK97}. Likewise, data
on the electromagnetic form factor of $\eta$ is consistent with
$|f_\eta^{(c)}|$ being small \cite{FK97}. With our estimate of 
$|f_\eta^{(c)}|$, 
we find that this charm-induced contribution does not dominate the
matrix element for $B^\pm \to \eta^\prime K^\pm$; the penguins play a
more important role numerically in this decay.

The branching ratio ${\cal B} ( B^\pm \to \eta^\prime K^\pm )$ 
as well as those of the related ones  $B^\pm \to \eta^\prime K^{\ast \pm }$,
 $B^\pm \to \eta K^\pm$, and
$B^\pm \to \eta K^{\ast \pm }$, depend upon the
interference of the amplitudes arising from the chain  $b \to s 
(\bar{c} c) \to s(\eta^\prime, \eta)$, and the ones arising from
calculating the matrix elements of the rest 
of the operators. Concentrating on
the decay $ B^\pm \to \eta^\prime K^\pm $, we note that the sign of the 
term involving the $b \to s
(\bar{c} c) \to s(\eta^\prime, \eta)$ in the full amplitude
 is not determined {\it a priori}. 
 Since the solutions with constructive or destructive
interference terms are both logical possibilities, we have estimated 
${\cal B} ( B^\pm \to \eta^\prime K^\pm )$
for both cases, with the positive-$f_{\eta^\prime}^{(c)}$ solution 
yielding marginally larger rate.
 However, more importantly, we find that the rate in this 
decay (and in some others) depends significantly on the parameter
$\xi$. Hence, to make absolute predictions, the 
phenomenological value of this parameter has to be determined. We study
a number of measured $B \to h_1 h_2$ decays to estimate a 
range for $\xi$ which, given the present experimental errors and 
theoretical accuracy of our approach estimated by us as a factor 2 in 
rates, is 
understandably not very precise at this stage. The range $0 \leq \xi \leq 
0.5$ is consistent with data.

 This paper is organized as follows. In section 2, we review the effective
Hamiltonian for the non-leptonic $B$ decays and calculate the matrix 
elements of the operators at the quark level in the NLL precision.
In section 3, we formulate our
factorization ansatz to calculate the hadronic matrix elements in the 
two-body decays $ B \to h_1 h_2$.
 The matrix elements for various decay 
modes of interest are also detailed here, together with a brief review of
the mixing formalism for the $\eta$ - $\eta^\prime - \eta_c$ sector. Our 
estimates of the decay constants $f_{\eta^\prime}^{(c)}$
and $f_{\eta}^{(c)}$  relevant for the decays $B^\pm 
\to (\eta^\prime, \eta) (K^\pm, K^{\ast \pm})$ are also given here. 
Section 4
contains our numerical results. The input values for the various 
quantities (coupling constants, form factors, quark masses) are collected
here in several tables. We compare the branching ratios with the CLEO data 
varying the factorization-related parameter $\xi$ and  parameters of the 
CKM matrix \cite{CKM}. The potential impact of some of these decays on the 
CKM phenomenology
is illustrated in terms of the ratios of the branching ratios,
which are more reliably calculable. In particular,
the ratios $R_1 \equiv {\cal B}(B^0(\overline{B^0}) 
\to \pi^\mp K^\pm)/{\cal B}(B^\pm \to 
\pi^\pm K)$ and $R_2 \equiv {\cal B}(B^0 (\overline{B^0}) 
\to \pi^\mp h^\pm)/{\cal 
B}(B^\pm \to \pi^\pm K)$,
constrain the CKM-Wolfenstein parameter $\rho$
and $\eta$ \cite{Wolfenstein83}. The potential importance of $R_1$ in 
determining the angle
$\gamma$ has been emphasized by Fleischer and Mannel \cite{FM97-1,FM97-2}.
Interestingly, within the theoretical framework presented here,
the measured ratio $R_1= 0.65 \pm 0.40$ suggests (at $\pm 1 \sigma)$ that 
$\rho \ge 0$, which in turn implies $\gamma \leq 
90^\circ$, where $\gamma$ is one of the CP-violating angles of the unitarity
triangle. 
We also comment on the effect of an (assumed) enhanced coefficient of the 
chromomagnetic operator, $C_8(m_W)$, in  non-leptonic 
two-body $B$ decays.  This scenario has been discussed in the
context of new physics effects in $B$ decays \cite{Kagan95,CGG96,LNO97}. 
We find, using the decay $B^\pm \to K 
\pi^\pm$, that varying the ratio $C_8(m_W)/C_8(m_W)^{SM}$ in a large range
$(\pm 10)$ has no appreciable effect on the branching ratio within the 
present accuracy. Finally, we conclude with a summary in section 5. 
\section{Effective Hamiltonian for the non-leptonic decays
$B \to h_1 h_2$}
\label{sec:2}
\setcounter{equation}{0}

We write the effective Hamiltonian $H_{eff}$ 
for the $\Delta B=1$ transitions
as 
\be
\label{heff}
H_{eff}
= \frac{G_{F}}{\sqrt{2}} \, \left[ \vub \vuqst
\left (C_1 O_1^u + C_2 O_2^u \right) 
+ \vcb \vcqst
\left (C_1 O_1^c + C_2 O_2^c \right) -
\vtb \vtqst \, 
\sum_{i=3}^{8}
C_{i} \, O_i \right] \quad ,
\ee
where
$q=d,s$ and $C_i$ are the Wilson coefficients evaluated at the 
renormalization scale $\mu$; the current-current operators $O_1^{u,c}$ and
$O_2^{u,c}$ read 
\bea
\label{cc}
O_1^u &=& \left( \bar{u}_\a b_\a \right)_{V-A} \,
        \left( \bar{q}_\b u_\b \right)_{V-A} \quad \quad
O_1^c = \left( \bar{c}_\a b_\a \right)_{V-A} \,
        \left( \bar{q}_\b c_\b \right)_{V-A} \nonumber \\
O_2^u &=& \left( \bar{u}_\b b_\a \right)_{V-A} \,
        \left( \bar{q}_\a u_\b \right)_{V-A} \quad \quad
O_2^c = \left( \bar{c}_\b b_\a \right)_{V-A} \,
        \left( \bar{q}_\a c_\b \right)_{V-A} \quad , 
\eea
while the QCD penguin operators $O_3-O_6$ are
\bea
\label{peng}
O_3 &=& \left( \bar{q}_\a b_\a \right)_{V-A} \, \sum_{q'}
        \left( \bar{q}'_\b  q'_\b \right)_{V-A} \quad
O_5  =  \left( \bar{q}_\a b_\a \right)_{V-A} \, \sum_{q'}
        \left( \bar{q}'_\b  q'_\b \right)_{V+A} \quad \nonumber \\
O_4 &=& \left( \bar{q}_\b b_\a \right)_{V-A} \, \sum_{q'}
        \left( \bar{q}'_\a  q'_\b \right)_{V-A} \quad
O_6  =  \left( \bar{q}_\b b_\a \right)_{V-A} \, \sum_{q'}
        \left( \bar{q}'_\a  q'_\b \right)_{V+A} \quad . 
\eea
Finally, the dipole operators $O_7$ and $O_8$ read 
\bea
\label{o78}
O_7 &=& (e/8\p^{2}) \, m_b \, \bar{s} \, \sigma^{\m \n}
      \, (1+\g_5)   \,b
      \ F_{\m \n} \quad , \nonumber \\
O_8 &=& (g_s/8\p^{2}) \, m_b \, \bar{s}_{\a} \, \sigma^{\m \n}
      \, (1+\g_5)  \, (\l^A_{\a \b}/2) \,b_{\b}
      \ G^A_{\m \n} \quad .
\eea
Here $\alpha$ and $\beta$ are the $SU(3)$ color indices and $\l^A_{\a \b},
A=1,...,8$, are the Gell-Mann matrices. The subscripts $V \pm A$ represent
the chiral projections $1 \pm \gamma_5$. Thus,  
in eqs. (\ref{cc}) and (\ref{peng}) $(\bar{u}_\a b_\b)_{V-A}
=\bar{u}_\a \gamma^\mu (1-\gamma_5) \, b_\b$ etc.
In eq.~(\ref{o78}) $F_{\m \n}$ and $G^A_{\m \n}$
denote the photonic and the gluonic field strength tensor,
respectively.
We note that we neglect the effects of the operator $O_7$ in the
present analysis as well as 
the so-called electro-weak penguin (4-Fermi) operators
which we did not list explicitly. 
Likewise, the
effect of weak annihilation and exchange diagrams will be neglected here.
This is in line with the investigations reported in literature
\cite{GHLR95-1}. 
Working consistently to NLL precision, 
the coefficients $C_1$-$C_6$ are needed
in NLL precision, while it is sufficient to use the LL value for
$C_8$. The relevant Wilson coefficients to the desired accuracy are 
listed in table 1 for the two scales $\mu =5.0$ GeV and $\mu =2.5$ GeV,
where $C_7^{eff}=C_7-C_5/3 -C_6$ and $C_8^{eff}=C_8 + C_5$.

\begin{table}[htb]
\label{coeff}
\begin{center}
\begin{tabular}{| c | r | r | r | }
\hline
 $C_i(\mu)$ & $\mu=5.0$ GeV & $\mu=2.5$ GeV\\
\hline \hline
$C_1^{NLL}$ & $1.070$ & $1.117$ \\
$C_2^{NLL}$ & $-0.166$ & $-0.257$ \\
$C_3^{NLL}$ & $0.011$ & $0.017$ \\
$C_4^{NLL}$ & $-0.031$ & $-0.044$ \\
$C_5^{NNL}$ & $0.009$ & $0.011$ \\
$C_6^{NNL}$ & $-0.037$ & $-0.056$ \\
$C_7^{eff,LL}$ & $-0.303$ & $-0.338$ \\
$C_8^{eff,LL}$ & $-0.144$ & $-0.158$ \\
\hline
\end{tabular}
\end{center}
\caption{Wilson coefficients $C_i(\mu )$ at the renormalization 
scale 
$\mu =5.0$ GeV and $\mu = 2.5$ GeV in the NDR scheme. 
$C_1 - C_6$ are in NLL accuracy, while $C_7^{eff}$
and $C_8^{eff}$ are in LL precision. 
For $\a_s(\mu)$
(in the $\overline{MS}$ scheme) we used the
two-loop expression with 5 flavors and $\a_s^{\overline{MS}}(m_Z)=0.118$;
$m_t^{\overline{MS}}(m_t)=165$ GeV (equivalent to
$m_{t,pole}= 175 $ GeV)..}
\label{table1}
\end{table}

\subsection{Quark-level matrix elements}
\begin{figure}[htb]
\vspace{0.10in}
\centerline{
\epsfig{file=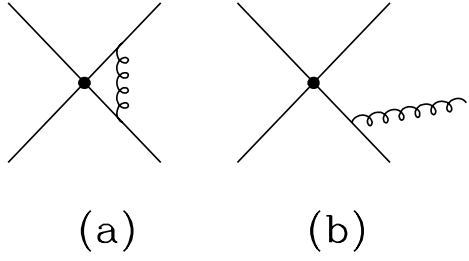,
height=1.5in,angle=0,clip=}
}
\vspace{0.08in}
\caption[]{(a) Vertex correction diagrams to the four-Fermi operators
$O_i$ $(i=1,...,6)$;\\ (b) Corresponding Bremsstrahlung corrections.
\label{vertexdiagram}}
\end{figure}

The  Wilson coefficients of the four-Fermi operators 
depend on the renormalization scale; in addition, in NLL precision,
they also depend on the renormalization scheme. 
These unphysical dependences 
are compensated in principle by a
corresponding scheme/scale dependence of the matrix elements 
of the operators. However, when using 
the factorization
ansatz for the hadronic matrix elements of the operators, these 
cancellations do not take place in practice, because the factorized
matrix elements of the operators are expressed in terms of decay constants
and form factors, and are as such scheme/scale independent.
To achieve this cancellation, we include perturbative QCD corrections to the
partonic matrix element before doing the factorization step.
We fully calculate the one-loop penguin-like diagrams in
Fig.~\ref{feynmandiagram}(a) and some
process independent parts (see below) of the vertex
correction diagrams   associated
with the four-Fermi operators, as  shown 
in Figs. \ref{vertexdiagram}(a). 
\begin{figure}[htb]
\vspace{0.10in}
\centerline{
\epsfig{file=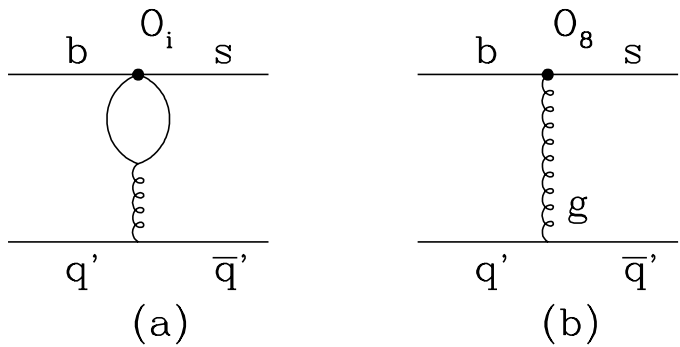,
height=2.0in,angle=0,clip=}
}
\vspace{0.08in}
\caption[]{(a) Penguin-type diagrams associated with the four-Fermi
operators $O_i$ ($i=1,...,6$); (b) Tree-level matrix element of the
chromomagnetic dipole operator $O_8$. 
\label{feynmandiagram}}
\end{figure}
These two classes of corrections are sufficient 
concerning the cancellation of
the scheme/scale dependences.
Furthermore, the contribution associated with the operator $O_8$,
where the gluon splits into a quark - antiquark pair, 
as shown in Fig.~\ref{feynmandiagram}(b) 
is of the same order in $\a_s$ as the corrections just mentioned and 
is therefore also taken into account in our analysis. 

As we use in this paper the Wilson coefficients obtained in the
naive dimensional scheme (NDR) with anti-commuting $\g_5$,
we also have to evaluate the various $O(\a_s)$ corrections  
in this scheme.   
These corrections can be absorbed
into effective
Wilson coefficients $C_i^{eff}$, which for a general $SU(N)$
color group can be written as
\bea
\label{ceff}
C_1^{eff} &=& C_1 + 
\frac{\a_s}{4\p} \, \left( r_V^T +
 \g_{V}^T \log \frac{m_b}{\mu}\right)_{1j} \, C_j  +\cdots  
\nonumber \\
C_2^{eff} &=& C_2 +  
\frac{\a_s}{4\p} \, \left( r_V^T +
 \g_V^T \log \frac{m_b}{\mu} \right)_{2j} \, C_j   +\cdots
\nonumber \\
C_3^{eff} &=& C_3 -\frac{1}{2N} \frac{\a_s}{4\p} \, (C_t + C_p + C_g) +
\frac{\a_s}{4\p} \, \left(  r_V^T +
\g_V^T \log \frac{m_b}{\mu}\right)_{3j} \, C_j +\cdots   
\nonumber \\
C_4^{eff} &=& C_4 +\frac{1}{2} \frac{\a_s}{4\p} \, (C_t + C_p + C_g) +
\frac{\a_s}{4\p} \, \left( r_V^T +
\g_V^T \log \frac{m_b}{\mu}\right)_{4j} \, C_j   +\cdots
\nonumber \\
C_5^{eff} &=& C_5 -\frac{1}{2N} \frac{\a_s}{4\p} \, (C_t + C_p + C_g) +
\frac{\a_s}{4\p} \, \left(  r_V^T +
\g_V^T \log \frac{m_b}{\mu}\right)_{5j} \, C_j   +\cdots
\nonumber \\
C_6^{eff} &=& C_6 +\frac{1}{2} \frac{\a_s}{4\p} \, (C_t + C_p + C_g) +
\frac{\a_s}{4\p} \, \left(  r_V^T +
 \g_V^T \log \frac{m_b}{\mu} \right)_{6j} \, C_j   +\cdots
\quad .
\eea
We have separated the contributions $C_t$, $C_p$, and $C_g$
arising from the penguin-type diagrams of 
the current-current operators $O_{1,2}$, the penguin-type diagrams of
the operators
$O_3$-$O_6$, and the tree-level diagram of the 
dipole operator $O_8$, respectively. 
The process-independent contributions 
from the vertex-type diagrams are contained in the matrices
$r_V$ and $\g_V$. Here 
$\g_V$ is that part of the anomalous matrix which is due
to the vertex  (and self-energy) corrections. This part can 
be easily extracted from $\hat\g^{(0)}$ in ref. \cite{Burasetal92}:
\be
\label{gammamatrix}
\g_V = \left(
\begin{array}{cccccc}
-2 & 6 & 0 & 0 & 0 & 0 \\
6 & -2 & 0 & 0 & 0 & 0 \\
0 & 0  &-2 & 6 & 0 & 0 \\
0 & 0  & 6            & -2           & 0 & 0 \\
0 & 0  & 0            & 0            & 2 & -6 \\
0 & 0  & 0            & 0            & 0 & -16 \\
\end{array}
\right) \quad .
\ee
The matrix $r_V$ contains constant, i.e.,  momentum-independent parts
associated with the vertex diagrams. This matrix
can be extracted from the matrix $\hat{r}$ defined
in eq.~(2.12) (and given explicitly in eq.~(4.6) 
in ref. \cite{Burasetal92}):
\be
\label{rmatrix}
r_V = \left(
\begin{array}{cccccc}
\frac{7}{3} & -7 & 0 & 0 & 0 & 0 \\
-7 & \frac{7}{3} & 0 & 0 & 0 & 0 \\
0 & 0  & \frac{63}{27} & -\frac{63}{9} & 0 & 0 \\
0 & 0  & -7            & \frac{7}{3}          & 0 & 0 \\
0 & 0  & 0            & 0            & -\frac{1}{3} & 1 \\
0 & 0  & 0            & 0            & -3 & \frac{35}{3} \\
\end{array}
\right) \quad .
\ee
Note that the $\mu$ dependence and the scheme dependence
of the vertex correction diagrams
are fully taken into account in eq.~(\ref{ceff}) by the terms
involving the matrices $\g_V$ and $r_V$, respectively.
There are, however, still scheme-independent, process-specific terms
omitted as indicated by the ellipses.
When calculating inclusive quantities, such as the
semileptonic branching ratios and $B$-hadron lifetimes, 
it is straightforward how to take these
corrections into account.
The virtual corrections are infrared divergent on their own, 
but together with the Bremsstrahlung
contributions in Fig.~\ref{vertexdiagram}(b), 
they lead to a finite and well defined $O(\a_s)$ correction, which is
found to be small. However,
 it is less obvious how to include them in exclusive two-body 
decays. The point is that the division of the final states
with and without the extra gluon is ambiguous and can be meaningfully
defined only with a cut-off. 
As such a separation into virtual corrections
 and soft gluon Bremsstrahlung contributions
is arbitrary anyhow, we only take into account the terms involving 
$\g_V$ and $r_V$.
The explicit $O(\a_s)$
contributions which emerge from the penguin operators involving a $q\bar{q}$
pair in the loop are infrared finite on their own and hence do not require
a cut-off.

 The quantities $C_t$, $C_p$, and $C_g$ given 
by the diagrams shown in Fig.~\ref{feynmandiagram}
read in  the NDR scheme
(after $\overline{MS}$ renormalization) 
\be
\label{ct}
C_t = - C_1 \, \left[ 
\frac{V_{cb}V_{cq}^*}{V_{tb} V_{tq}^*} \, \tilde{C}_t(m_c) + 
\frac{V_{ub}V_{uq}^*}{V_{tb} V_{tq}^*} \, \tilde{C}_t(m_u) \right] \quad
, \quad 
\tilde{C_t}(m) =
\frac{2}{3} + \frac{2}{3} \log \frac{m^2}{\mu^2}
- \Delta F_1 \left( \frac{q^2}{m^2} \right)  \quad ,
\ee  
\bea
\label{cp}
C_p &=& C_3 \, \left[ \frac{4}{3} 
+ \frac{2}{3} \log \frac{m_q^2}{\mu^2}
+ \frac{2}{3} \log \frac{m_b^2}{\mu^2}
- \Delta F_1 \left( \frac{q^2}{m_q^2} \right) 
- \Delta F_1 \left( \frac{q^2}{m_b^2} \right) 
\right] \quad , \nonumber \\
&& + (C_4 + C_6) \, \sum_{i=u,d,s,c,b} \, \left[ \frac{2}{3} \, \log
\frac{m_i^2}{\mu^2} - \Delta F_1 \left( \frac{q^2}{m_i^2} \right) \,
\right] \quad ,
\eea  
\be
\label{cg}
C_g = - \frac{2 m_b}{\sqrt{\bra q^2 \ket}} \, C_8^{eff} \quad ,
\ee
with $C_8^{eff}=C_8+C_5$. The function $\Delta F_1(z)$ is 
defined as
\be
\label{deltaf1}
\Delta F_1(z) = -4 \, \int_0^1 dx \, x(1-x) \, \log \left[
1-z \, x(1-x) - i \epsilon \right] \quad .
\ee

Two remarks are in order here. First, 
our expressions for $C_i^{eff}$
in eq.~(\ref{ceff}) are written in terms of the 
Wilson coefficients in the NDR scheme. Analogous expressions (but with 
$r_V=0$, $\g_V=0$ and $C_g=0$) 
have been obtained earlier in the literature \cite{KPS94}. Comparing
the expressions given here with the ones in 
\cite{KPS94}, where the corresponding quantities $c_3^{eff},...,c_6^{eff}$  
are expressed
in terms of the so-called renormalization-scheme-independent Wilson 
coefficients $\bar{c}_i$ introduced in ref. \cite{Burasetal92}, one
notices that the constant terms appearing explicitly in
 $C_t$ and $C_p$ in the two papers are 
different.
 As the scheme dependence
cancels automatically when including the one-loop matrix elements
discussed above,
we prefer to work with the Wilson coefficients in the NDR
scheme.

 Second, we have to explain the assumption  which
allows us to absorb the tree-level diagram 
$b \to s g \to s \bar{q}' q'$ 
associated with the operator $O_8$ into the contribution $C_g$
appearing in the expressions for $C_i^{eff}$. It is straightforward
to write down the matrix element
\be
\label{matel}
\bra s \bar{q}' q' | O_8 | b \ket = - \frac{\a_s}{\p} \, 
\frac{m_b}{q^2} \, \left( \bar{s}_\a \g_\mu \qsl 
(1+\g_5) \, \frac{\l^A_{\a\b}}{2} b_\b \right) \,
\left(
\bar{q}'_\g \g^\mu \frac{\l^A_{\g \d}}{2} \, q'_\d 
\right) \quad ,
\ee
where $q$ is the momentum transferred by the gluon to the 
($q'$,$\bar{q}'$)-pair.
In the factorization model to be described below, $q'$ and $\bar{q}'$
cannot go into the same meson in the process $B \to h_1 h_2$
due to color, i.e., $q'$ goes into $h_1$, while $\bar{q}'$ goes into
$h_2$ or vice versa (see Fig.~\ref{feynmandiagram}(b)).
The quantities $C_t,~C_p$ and $C_g$ depend on the momentum 
$q$. Since we are interested here only in two-body decays, we assume 
for simplicity, 
that the three-momenta of $q'$ and $\bar{q}'$ are equal in magnitude
but opposite in direction 
in the rest frame of the $b$-quark.
The momentum transfer $q$ is then proportional to
$p_b$, i.e.,
\be
\label{kinematik}
q^\mu = \sqrt{\bra q^2 \ket} \, \frac{p_b^\mu}{m_b} \quad ,  
\ee
where $\bra q^2 \ket$ is an averaged value of $q^2$.
Inserting (\ref{kinematik}) into eq.~(\ref{matel}) and using
the equations of motion, the expression for $C_g$ in
eq.~(\ref{cg}) is readily obtained. To be consistent, we should
also replace $q^2$ by $\bra q^2\ket $ in the expressions
for $C_t$ and $C_p$ in eqs. (\ref{ct}) and (\ref{cp}), 
respectively. To estimate the theoretical uncertainty introduced thereby, 
 we treat $\bra q^2 \ket$ as a parameter  which varies 
in the range $m_b^2/4 \le \bra q^2 \ket \le m_b^2/2$, following the
prescriptions in literature \cite{DT90,SW91}.

To summarize: 
The various $O(\a_s)$ corrections have been absorbed into effective
Wilson coefficients $C_i^{eff}$ ($i=1,...,6$); these coefficients
are  scheme independent and
the term $\sim \a_s \log \mu$, which dominates the scale dependence
of the original Wilson coefficients $C_i$ 
and the one-loop matrix elements, is absent in $C_i^{eff}$.
What remains to be done is to estimate the hadronic matrix elements
$\bra h_1 h_2 |C_i^{eff} \, O_i|B \ket$ for $i=1,...,6$.
The numerical values of the quantities $C_i^{eff}$ are given
in table 5 in section 4.1.4.  
 
\section{Factorization ansatz for the matrix elements 
in $B \to h_1 h_2$}
\begin{figure}[htb]
\vspace{0.10in}
\centerline{
\epsfig{file=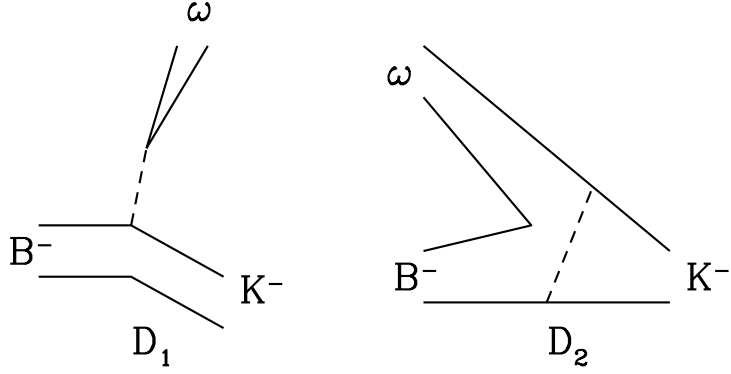,
height=2.0in,angle=0,clip=}
}
\vspace{0.08in}
\caption[]{$D_1$ and $D_2$ are the two diagrams contributing in the
factorization approximation. See text.
\label{factordiagram}}
\end{figure}

\setcounter{equation}{0}
We have now to work out the hadronic matrix elements
of the operators $O_i$ $(i=1,...,6)$ for the processes
of interest. We use the factorization approximation, which we briefly
explain for a specific example.
Consider the matrix element due to the $u$-quark contribution of
the operator $O_5$ for the process $B^- \to K^- \omega$, i.e.,
\be
\bra K^- \omega |O_5^{(u)} | B^-\ket \quad,
\quad O_5^{(u)} = \left(\bar{s} \g_\mu \, (1-\g_5) b  \right) \,
                  \left(\bar{u} \g^\mu \, (1+\g_5) u  \right) \quad.
\ee 
There are  two contributing diagrams
$D_1$ and $D_2$ shown in Fig.~\ref{factordiagram}.
The factorization approximation for $D_1$ is readily obtained:
\be
\label{d1}
D_1 = \bra \omega | \bar{u} \g^\mu \, (1+\g_5) \, u|0 \ket \,
\bra K^- | \bar{s} \g_\mu \, (1-\g_5) \, b |B^-\ket=
 \bra \omega |\bar{u}u_-|0 \ket \, \bra K^- |\bar{s} b_- |B^- \ket 
\quad ,
\ee
where here and in the following  the short-hand notation
$\bar{q}q'_-$ stands for
\be
\bar{q}q'_- = \bar{q} \, \g_\mu \, (1-\g_5) \, q' \quad .
\ee
To get $D_2$ in the factorization approximation, we first write
the operator $O_5^{(u)}$ in its Fierzed form
\bea
\label{o5fierz}
O_5^{(u)} &=& -2 \, \left( \bar{u}_\b (1-\g_5) b_\a \right)
\left( \bar{s}_\a (1+\g_5) u_\b \right) \nonumber \\
&& =-2 \, \left[ \frac{1}{N} \, \left( \bar{u} (1-\g_5) b \right)
\left( \bar{s} (1+\g_5) u \right) +
\frac{1}{2}  \, \left( \bar{u} (1-\g_5) \l b \right)
\left( \bar{s} (1+\g_5) \l u \right) \right] \quad ,
\eea
where $\l$ denotes a color matrix. 
Only the first term in the square bracket in eq.~(\ref{o5fierz})
(being color singlet-singlet) contributes in the factorization 
approximation. One gets \be
\label{d2}
D_2 = - \, \frac{2}{N} \, \bra \omega | \bar{u} (1- \g_5) b |B^-\ket \,
 \bra K^- | \bar{s} (1 + \g_5) u |0 \ket \quad;
\ee
using the Dirac equation, we can write $D_2$ as
\be
\label{d2fin}
D_2 = -\frac{2}{N} \, \frac{m_K^2}{(m_s+m_u)\,(m_b+m_u)} \,  
\, \bra \omega | \bar{u} b_- |B^-\ket \,
 \bra K^- | \bar{s}  u_- |0 \ket \quad.
\ee
Doing analogous manipulations,
the complete matrix element $M$ 
for $B^- \to \omega K^-$, defined as
\be
\label{hamfin}
M = \bra \omega K^-|H_{eff}| B^-\ket \quad; \quad
H_{eff} =\frac{G_F}{\sqrt{2}} \, \left[ V_{ub} V_{us}^* \, 
\left(C_1^{eff} \, O_1^u + C_2^{eff} \, O_2^u \right) \,
-V_{tb} V_{ts}^* \, \sum_{i=3}^6 \, C_i^{eff} O_i \right] \quad ,
\ee
is then easily
obtained. One gets
\bea
\label{proc0}
M &=& \frac{G_F}{\sqrt{2}} \left\{ \, 
V_{ub} \, V_{us}^* \, 
\left( a_1 \, 
\bra K^-|\bar{s} u_-|0\ket  \,
\bra \omega|\bar{u} b_-|B^- \ket
+ a_2 \, 
\bra K^-|\bar{s} b_-|B^-\ket \,
\bra \omega|\bar{u} u_-|0 \ket \right) \right. \nonumber \\
&& - V_{tb} V_{ts}^* \, 
\left( \left(a_4 - \frac{2 a_6 m_K^2}{(m_s+m_u)\, (m_b+m_u)} \right) \, 
\bra K^-|\bar{s} u_-|0\ket  \,
\bra \omega|\bar{u} b_-|B^- \ket \right. \nonumber \\
&& \left. \left. + 2 (a_3+ a_5) \, 
\bra K^-|\bar{s} b_-|B^-\ket \,
\bra \omega|\bar{u} u_-|0 \ket \right) \right\} \quad .
\eea
The quantities $a_i$ $(i=1,...,6)$
are the following combinations 
of the effective Wilson coefficients in eq.~(\ref{ceff}):
\be
\label{ai}
a_{2i-1} = C^{eff}_{2i-1} + \frac{1}{N} \, C^{eff}_{2i} \quad , \quad
a_{2i} = C^{eff}_{2i} + \frac{1}{N} \, C^{eff}_{2i-1} \quad , \quad
i=1,2,3 \quad .
\ee
The explicit $1/N$ terms in eq.~(\ref{ai}) 
are always accompanied by an octet-octet contribution;
this can be seen explicitly in eq.~(\ref{o5fierz}).
As one discards this octet-octet contribution in the
 factorization approximation, one usually  replaces $1/N$ by $\xi$
and treats $\xi$ as a free parameter with the hope to compensate
phenomenologically for the
omitted octet-octet contribution in terms of a rescaled value of $\xi$. 
Note, however, that the $1/N$ factors appearing explicitly in 
the (perturbative) expressions  
for the effective Wilson coefficients in eq.~(\ref{ceff}) are not replaced
by $\xi$ in our work in contrast to \cite{KPS94}, where also these
$1/N$ factors were replaced by $\xi$. We think that replacing also
these $1/N$ terms by $\xi$  destroys the scheme independence
of the effective Wilson coefficients.  

 It is worth pointing out that the factorization Ansatz just discussed is
the simplest one. Also, it is implicitly assumed that the relative strong
phases (such as the ones arising from the final state interactions but
also due to the non-perturbative contributions to the charm-penguins),
contributing to the different $a_i$'s  are
small. Of course, it does not mean that the strong interaction phases are
assumed absent. The ones generated by the next-to-leading order 
perturbative QCD contributions from the charm penguins are taken into
account. It remains to be seen if the non-perturbative phases from the
competing tree and penguin contributions in the processes discussed here are 
indeed small. 

Finally, before giving the matrix elements for the various
exclusive two-body decays, we discuss the parametrization of
the decay constants and form factors which appear in the factorized
form of the hadronic matrix elements.
The form factors are parametrized as
\be
\bra P(p')|V_\mu|B(p)\ket = [(p+p')_\mu - 
\frac{m_B^2-m_P^2}{q^2} \, q_\mu \,] \, F_1(q^2) + 
\frac{m_B^2-m_P^2}{q^2} \, q_\mu \, F_0(q^2)
\ee 
\bea
&& \bra V(\epsilon,p')| ( V_\mu-A_\mu)| B(p)\ket = 
\frac{2}{m_B+m_V} \, i \, \e_{\m\n\a\b} \epsilon^{*\n} \, p^\a \, p'^\b \,
 V(q^2) \nonumber \\
&& - (m_B+m_V) \, \left[ \epsilon^*_\mu - 
\frac{\epsilon^* \cdot q}{q^2} \, q_\mu \right] \, A_1(q^2) +
\nonumber \\
&& \frac{\epsilon^* \cdot q}{m_B + m_V} \, \left[
(p+p')_\mu - \frac{m_B^2-m_V^2}{q^2} \, q_\mu \, \right] \,
A_2(q^2) - 
\epsilon^* \cdot q \, \frac{2 m_V}{q^2} \, q_\mu \, A_0(q^2) \quad ,
\eea
where $P(V)$ is a pseudoscalar(vector) meson, $q=p-p'$,
\be
A_0(0) = \frac{m_B+m_V}{2 m_V} \, A_1(0) -
         \frac{m_B-m_V}{2 m_V} \, A_2(0) 
\ee
and $F_1(0) = F_0(0)$. The decay constants $f_P$ and $f_V$
are defined as
\be
\label{decay}
\bra 0 |A_\mu|P(p) \ket = i \, f_P \, p_\mu \quad, \quad
\bra 0 |V_\mu|V(\epsilon,p) \ket = i \, f_V \, m_V \,
\epsilon_\mu \quad.
\ee
With these definitions we are in a position to write down the formulas for 
the matrix elements for the two body decays. They are given below 
explicitly for the four generic decay modes: $B \to \pi \pi$, $B\to K \pi$,
$B^\pm \to K^\pm \omega, \pi^\pm \omega$ and $ B^\pm \to (K^\pm,K^{\ast \pm})
(\eta,\eta^\prime)$, which are also the ones we calculate numerically
in the next section. However, the formalism given here is general and 
applicable to all two-body $B$-decays of the type $B \to PP$, $B \to 
PV$, and $B \to VV$. 
\subsection{$B \to \pi \pi$}
In this section we discuss the processes 
$B^\pm \to \pi^\pm \pi^0 $, $B^0 (\overline{B^0}) \to \pi^\pm \pi^\mp$
and $B^0 (\overline{B^0}) \to \pi^0 \pi^0$.
\subsubsection{$B^\pm \to \pi^\pm \pi^0 $} 
The matrix element $M$ for $B^- \to  \pi^- \pi^0$ involves the operators 
$O_1^u$ and $O_2^u$ and reads (neglecting SU(2) breaking effects) 
\be
\label{proc5}
M = \frac{G_F}{\sqrt{2}} \,  V_{ub} V_{ud}^* \, (a_1 +a_2) \,
 \bra \pi^- | \bar{d} \, u_-|0 \ket \, 
            \bra \pi^0 | \bar{u} \, b_-|B^- \ket \quad , 
\ee
with
\be
\label{proc5a}
\bra \pi^-|\bar{d} u_- |0 \ket \, \bra \pi^0|\bar{u} b_-|B^- \ket =
i f_\pi \, (m_B^2-m_\pi^2) \, F_0^{B \to \pi^0}(m_\pi^2) \quad .
\ee
The branching ratio ${\cal B}(B^- \to \pi^- \pi^0)$ is then given by
the expression
\be
{\cal B}(B^- \to \pi^- \pi^0) = \tau_B \, \frac{1}{8\p} \, |M|^2 \, 
\frac{|p|}{m_B^2} \quad ,
\ee
where $\tau_B$ is the lifetime of the $B^0$-meson
and $|p|$ is the absolute value of the 3-momentum of the $\pi^-$
(or the $\pi^0$) in the rest frame of the $B^0$ meson. This expression
for the branching ratio
holds for other two-body decays being discussed with obvious changes of
the indicated quantities. Hence, we shall give subsequently only the
matrix elements $M$. Also, we shall give only the amplitudes for the
decays of $B^-$ and $\overline{B^0}$, and the
matrix elements for the charge-conjugate processes are then obtained
by complex conjugating the CKM factors. Since we are not addressing the
question of CP-violation in this paper, all decay rates given later  are 
to be 
interpreted in terms of the averaged branching ratios. Thus, for example, the
branching ratio ${\cal B}(B^\pm \to \pi^\pm \pi^0)$ is  defined
as
\be
{\cal B}(B^\pm \to \pi^\pm \pi^0) = \frac{1}{2} \, \left(
{\cal B}(B^- \to \pi^- \pi^0) + {\cal B}(B^+ \to \pi^+ \pi^0) \right) \quad .
\ee
\subsubsection{$B^0 (\overline{B^0}) \to \pi^+ \pi^- $}
The matrix element $M$ for $\overline{B^0} \to  \pi^- \pi^+$ reads 
\bea
\label{proc5c}
M &=& \frac{G_F}{\sqrt{2}} \,  \left[ V_{ub} V_{ud}^* \, a_1 \,
-V_{tb}V_{td}^* \, \left( a_4 +\frac{2 \, a_6 \,
 m_\pi^2}{(m_b-m_u)\,(m_u+m_d)} \right) \right] \times \nonumber \\
&& \bra \pi^- | \bar{d} \, u_-|0 \ket \, 
            \bra \pi^+ | \bar{u} \, b_-|\bar{B}^0 \ket \quad , 
\eea
with
\be
\label{proc5d}
\bra \pi^-|\bar{d} u_- |0 \ket \, \bra \pi^+|\bar{u} b_-|\bar{B}^0 \ket =
i f_\pi \, (m_B^2-m_\pi^2) \, F_0^{B \to \pi^-}(m_\pi^2) \quad .
\ee
\subsubsection{$B^0 (\overline{B^0}) \to \pi^0 \pi^0 $}
The matrix element $M$ for $\overline{B^0} \to  \pi^0 \pi^0$ reads 
\bea
\label{proc5e}
M &=& \frac{G_F}{\sqrt{2}} \,  \left[ V_{ub} V_{ud}^* \, a_2 \,
+V_{tb}V_{td}^* \, \left( a_4 + \frac{ \, a_6 \,
 m_\pi^2}{m_d \, (m_b-m_d)} \right) \right] \times \nonumber \\
&&  2 \, \bra \pi^0 | \bar{u} \, u_-|0 \ket \, 
            \bra \pi^0 | \bar{d} \, b_-|\bar{B}^0 \ket \quad , 
\eea
with
\be
\label{proc5f}
\bra \pi^0|\bar{u} u_- |0 \ket \, \bra \pi^0|\bar{d} b_-|\bar{B}^0 \ket =
i \frac{f_\pi}{\sqrt{2}} \, (m_B^2-m_\pi^2) \, 
F_0^{B \to \pi^0}(m_\pi^2) \quad .
\ee
When calculating the decay width, we have to take into account an extra
factor $1/2$ due to the two identical particles in the final state.
\subsection{$B \to K \pi$, $B \to K K$}
\subsubsection{$B^\pm \to K \pi^\pm$}
The matrix element $M$ for $B^- \to \pi^- \bar{K}^0 $ reads 
\bea
\label{proc3}
M &=& -\frac{G_F}{\sqrt{2}}
\, V_{tb} V_{ts}^* \, \left[a_4 +  \frac{2a_6 \, 
m_K^2}{(m_b-m_d)\,(m_s+m_d)}
 \right] \,
 \bra \pi^- | \bar{d} \, b_-|B^- \ket \, 
            \bra \bar{K}^0 | \bar{s} \, d_-|0 \ket 
\eea
with
\be
\label{proc3a}
 \bra \pi^- | \bar{d} \, b_-|B^- \ket \, 
            \bra \bar{K}^0 | \bar{s} \, d_-|0 \ket =
i \, f_{K} \, (m_B^2-m_\pi^2) \, 
F_0^{B \to \pi^-}(m_K^2) \quad . 
\ee
\subsubsection{$B^0(\overline{B^0}) \to  K^\pm \pi^\mp$}
The matrix element $M$ for $\overline{B^0} \to  \pi^+ K^-$ reads 
\bea
\label{proc4}
M &=& \frac{G_F}{\sqrt{2}} \, \left[ V_{ub} V_{us}^* \, a_1
- V_{tb} V_{ts}^* \, \left( a_4+  
2 a_6 \, \frac{m_K^2}{(m_b-m_u)\,(m_s+m_u))} \right) \, \right]
\times \nonumber \\ 
&& \bra \pi^+ | \bar{u} \, b_-|\bar{B}^0 \ket \, 
            \bra K^- | \bar{s} \, u_-|0 \ket 
\eea
with
\be
\label{proc4a}
 \bra \pi^+ | \bar{u} \, b_-|\bar{B}^0 \ket \, 
            \bra K^- | \bar{s} \, u_-|0 \ket =
i \, f_{K} \, (m_B^2-m_\pi^2) \,
F_0^{B \to \pi^-}(m_K^2) \quad . 
\ee 

\subsubsection{$B^\pm \to K^\pm \pi^0$}
The matrix element $M$ for $B^- \to K^- \pi^0 $ is given by
\bea
\label{proc5b}
M &=& \frac{G_F}{\sqrt{2}} \, \left[ V_{ub} V_{us}^* \, 
\left( a_1 + a_2 \, \frac{f_\pi}{\sqrt{2} f_K} \,
\frac{m_B^2-m_K^2}{m_B^2-m_\pi^2} \, 
\frac{F_0^{B \to K^-}(m_\pi^2)}{F_0^{B \to \pi^0}(m_K^2)} \right) 
\right. \nonumber \\ 
&& \left. - V_{tb} V_{ts}^* \, \left( a_4+  
2 a_6 \, \frac{m_K^2}{(m_b-m_u)\,(m_s+m_u))} \right) \, \right] \, 
\bra \pi^0 | \bar{u} \, b_-|B^- \ket \, 
            \bra K^- | \bar{s} \, u_-|0 \ket \quad .
\eea
with
\label{proc5c3}
\be
\bra K^-|\bar{s} u_- |0 \ket \, \bra \pi^0|\bar{u} b_-|B^- \ket =
i f_K \, (m_B^2-m_\pi^2) \, F_0^{B \to \pi^0}(m_K^2) \quad ,
\ee
\subsubsection{$B^\pm \to K^0 K^\pm $}
The matrix element $M$ for $B^- \to K^0 K^- $ is given by
\bea
\label{proc5b3}
M &=& - \frac{G_F}{\sqrt{2}} \,  V_{tb} V_{td}^* \, 
 \left( a_4+  
2 a_6 \, \frac{m_K^2}{(m_b-m_s)\,(m_s+m_d))} \right) \,  
\bra K^- | \bar{s} \, b_-|B^- \ket \, 
            \bra K^0 | \bar{d} \, s_-|0 \ket \quad , 
\eea
with
\label{proc5c23}
\be
\bra K^0|\bar{d} s_- |0 \ket \, \bra K^-|\bar{s} b_-|B^- \ket =
i f_K \, (m_B^2-m_K^2) \, F_0^{B \to K}(m_K^2) \quad .
\ee
\subsection{$B^\pm \to K^\pm \omega, \pi^\pm \omega$}
\subsubsection{$B^\pm \to K^\pm \omega$}
The matrix element $M$ for $B^- \to  K^- \omega$ reads 
\bea
\label{proc1}
M &=& \frac{G_F}{\sqrt{2}} \, \left\{ V_{ub} V_{us}^* \, \left[
a_1 + a_2  \, 
\frac{F_1^{B \to K^-}(m_\omega^2)}{A_0^{B \to \omega}(m_K^2)} \,
\frac{f_\omega}{\sqrt{2} f_K} \right] 
- V_{tb} V_{ts}^* \, \left[2 (a_3 + a_5) 
\, 
\frac{F_1^{B \to K^-}(m_\omega^2)}{A_0^{B \to \omega}(m_K^2)} \,
\frac{f_\omega}{\sqrt{2} f_K} \right. \right. \nonumber \\ 
&& \left. \left. + a_4 - \frac{2a_6 \, 
m_K^2}{(m_b+m_u)\,(m_s+m_u)}
 \right] \,
\right\} \, \bra \omega | \bar{u} \, b_-|B^- \ket \, 
            \bra K^- | \bar{s} \, u_-|0 \ket 
\eea
with
\be
\label{proc1a}
 \bra \omega | \bar{u} \, b_-|B^- \ket \, 
            \bra K^- | \bar{s} \, u_-|0 \ket =
-i \, f_{K} \, 2 \, m_\omega \, (p_B \cdot \epsilon^*_\omega ) \, 
A_0^{B \to \omega}(m_K^2) \quad . 
\ee 
There is only one non-vanishing helicity amplitude.
In the rest frame of the decaying $B$ meson only longitudinally
polarized $\omega$'s are produced. $p_B \cdot \epsilon^*_\omega$
is then given by
\be
p_B \cdot \epsilon^*_\omega = \frac{m_B}{m_\omega} \, |p| \quad ,
\ee
where $|p|$ is the absolute value of the 3-momentum of the
$\omega$ (or the $K^-$) in the $B$ rest frame.
\subsubsection{$B^\pm \to \pi^\pm \omega$}
The matrix element $M$ for $B^- \to  \pi^- \omega$ reads 
\bea
\label{proc2}
M &=& \frac{G_F}{\sqrt{2}} \, \left\{ V_{ub} V_{ud}^* \, \left[
a_1 + a_2  \, 
\frac{F_1^{B \to \pi^-}(m_\omega^2)}{A_0^{B \to \omega}(m_\pi^2)} \,
\frac{f_\omega}{\sqrt{2} f_\pi} \right] 
- V_{tb} V_{td}^* \, \left[(2 a_3 + a_4+ 2a_5) 
\, 
\frac{F_1^{B \to \pi^-}(m_\omega^2)}{A_0^{B \to \omega}(m_\pi^2)} \,
\frac{f_\omega}{\sqrt{2} f_\pi} \right. \right. \nonumber \\ 
&& \left. \left. + a_4 - \frac{2a_6 \, 
m_\pi^2}{(m_b+m_u)\,(m_d+m_u)}
 \right] \,
\right\} \, \bra \omega | \bar{u} \, b_-|B^- \ket \, 
            \bra \pi^- | \bar{d} \, u_-|0 \ket 
\eea
with
\be
\label{proc2a}
 \bra \omega | \bar{u} \, b_-|B^- \ket \, 
            \bra \pi^- | \bar{d} \, u_-|0 \ket =
-i \, f_{\pi} \, 2 \, m_\omega \, (p_B \cdot \epsilon^*_\omega ) \, 
A_0^{B \to \omega}(m_\pi^2) \quad . 
\ee 
\subsection{Mixing in the $\eta-\eta'-\eta_c$ system and
the decays $B^\pm \to K^{\pm} \eta^{(')},K^{* \pm} \eta^{(')}$}

Before we write the matrix elements for 
$B^\pm \to K^\pm \eta'$,
$B^\pm \to K^{* \pm } \eta'$,
$B^\pm \to K^\pm \eta$ and
$B^\pm \to K^{* \pm }\eta$
in the factorization approximation, we give a short
discussion about the $\eta$ - $\eta' - \eta_c$ system. Our main interest
for the decays mentioned above is to compute the hadronic matrix elements
$\langle 0 |(\bar{c}\gamma_\mu \gamma_5)|\eta \rangle$ and
$\langle 0 |(\bar{c}\gamma_\mu \gamma_5)|\eta' \rangle$.
 The conventional $(\eta,\eta')$-mixing formalism involves a
single mixing angle (called henceforth $\theta$) and it has been
argued that it provides a satisfactory description of the
decays involving $\eta$ and $\eta'$ \cite{GK87,VH97}. However, recently the
inadequacy of this mixing formalism has been pointed out in the context of
the $1/N_c$-improved chiral $U(3) \otimes U(3)$ perturbation theory
\cite{Leutwyler97,HSLT97}. Instead, a formalism which involves two mixing 
angles in 
the $SU(3)$-octet and -singlet sector (called henceforth $\theta_8$ and 
$\theta_0$) is proposed. Since, the $SU(3)$-singlet component
$|\eta_0 \rangle$ in general mixes with the $|c\bar{c}\rangle$ component, 
introducing
another angle (called $\theta_{c\bar{c}}$), we shall term the resulting 
mixing
formalisms as the two-angle (involving $\theta$ and $\theta_{c\bar{c}}$)
and three-angle (involving $\theta_0, ~\theta_8$ and $\theta_{c\bar{c}}$)
frameworks.

\subsubsection{$\eta$ - $\eta'-\eta_c$ system in the two-angle mixing 
formalism}
 Here, the physical $\eta$ and $\eta'$ states are considered as mixtures of
the $\eta_8$ and $\eta_0$ states \cite{GK87}:
\be
\label{mix}
|\eta \ket = \cos \theta | \eta_8 \ket - \sin \theta 
| \eta_0 \ket \quad , 
|\eta' \ket = \sin \theta | \eta_8 \ket + \cos \theta 
| \eta_0 \ket \quad , 
\ee 
where $\eta_8$ belongs to the SU(3) octet of the light pseudoscalar
(Goldstone) mesons, while $\eta_0$ is an SU(3) singlet. In the 
quark basis they are given by
\be
\label{quarkbasis}
| \eta_8 \ket = \frac{1}{\sqrt{6}} \, | u \bar{u} + d \bar{d} - 2
s \bar{s} \ket \quad , 
| \eta_0 \ket = \frac{1}{\sqrt{3}} \, | u \bar{u} + d \bar{d} +
s \bar{s} \ket \quad . 
\ee 
The mixing angle $\theta$ can be extracted from the measured
ratios \cite{GK87}
\be
\label{currenta}
\frac{\G(\eta \to \g\g)}{\G(\p^0 \to \g\g)} = 18 \,
\left( \frac{m_\eta}{m_\p} \right)^3 \, f_\pi^2
\, \left[ \frac{\cos \theta}{f_8} \, 
\frac{e_u^2 + e_d^2 - 2 e_s^2}{\sqrt{6}} -
\frac{\sin \theta}{f_0} \, 
\frac{e_u^2 + e_d^2 + e_s^2}{\sqrt{3}} \right]^2
\ee
and
\be
\label{currentb}
\frac{\G(\eta' \to \g\g)}{\G(\p^0 \to \g\g)} = 18 \,
\left( \frac{m_{\eta'}}{m_\p} \right)^3 \, f_\pi^2
\, \left[ \frac{\sin \theta}{f_8} \, 
\frac{e_u^2 + e_d^2 - 2 e_s^2}{\sqrt{6}} +
\frac{\cos \theta}{f_0} \, 
\frac{e_u^2 + e_d^2 + e_s^2}{\sqrt{3}} \right]^2 \quad ,
\ee
where $e_i$ are the quark charges, 
$f_\pi$, $f_8$ and $f_0$ are the decay constants of the pion,
the eighth component of the octet, and the singlet, respectively.
Using $f_8/f_\pi=1.34 \pm 0.03$ \cite{GaL85} and the measured decay 
widths \cite{PDG96}
$\Gamma(\pi^0 \to \g\g)=(7.7 \pm 0.55)$ eV,
$\Gamma(\eta \to \g\g)=(0.46 \pm 0.04)$ KeV,
$\Gamma(\eta' \to \g\g)=(4.26 \pm 0.19)$ KeV, one obtains
\be
\label{theta}
\theta = -21.3^o \pm 2.5^o \quad ; \quad
\frac{f_0}{f_\pi} = 1.09 \pm 0.05 \quad .
\ee

It has been suggested in the context of the radiative decays
$J/\psi \to \eta \gamma, ~\eta' \gamma$ that they can be enacted by
modelling them on the decay chain $J/\psi \to \eta_c \gamma \to (\eta, 
\eta') \gamma$, involving the Zweig-rule violating virtual transition
$\eta_c \to \eta$ and $\eta_c \to \eta'$. One can visualize these transitions
taking place via the two-gluon intermediate state. Since, only the
$SU(3)$-singlet component of the $\eta$ and $\eta'$ eigenstates is
involved, one obtains a relation for the ratio $R_{J/\psi} (\eta/\eta')$:
 \be
\label{RJPSI}
R_{J/\psi} (\eta/\eta') \equiv \frac{\G(J/\psi \to \eta' \g)}{\G(J/\psi \to
\eta\g)} = \left(\frac{k_{\eta'}}{k_{\eta}}\right)^3 \,
\frac{1}{\tan^2 \theta} \quad ,
\ee
where $k_{\eta'}$ and $k_{\eta}$ denote the 3-momenta
of $\eta'$ and $\eta$, respectively.
>From the measured ratio ($R_{J/\psi} (\eta/\eta')=5.0 \pm 0.8)$
\cite{PDG96},
a value of $|\theta|=21.9^o$ can be extracted. Thus, one gets a
consistent result from eqs. (\ref{theta}) and (\ref{RJPSI}).

 This Zweig-rule violating transition amplitude can also be
formulated by postulating that the physical $\eta'$ (as well as
the $\eta$) has a small effective charm component, which should enable 
us to enact
transitions of the type we are interested in, namely $b \to (c\bar{c}) s 
\to (\eta, \eta') s$. Thus,  there is
a contribution of the operators $O_{1,2}^c$ to the
decay amplitude for the processes
$B \to (K,K^*) (\eta', \eta)$, which can be modelled much the same way as the
decays $J/\psi \to (\eta, \eta') \gamma$.

 For $B^\pm \to K^\pm \eta^\prime$, for 
example, this yields in the factorization approximation
\be
\label{charmcontr}
M = - \frac{G_F}{\sqrt{2}} \, V_{cb} \, V_{cs}^* \, a_2 \,
\bra \eta'(q)|\bar{c} \g_\mu \g_5 c |0 \ket \, \bra
K(p')|\bar{s} \g^\mu b |B(p) \ket \quad .
\ee
The crucial quantity is the decay constant $ f^{(c)}_{\eta'}$
defined through the equation
\be
\bra \eta'(q)|\bar{c} \g_\mu \g_5 c |0\ket = -i f^{(c)}_{\eta'} \, q_\mu
\quad . 
\ee
The charm component comes in through the SU(3) 
singlet $|\eta_0 \ket$ which has a small charm admixture
characterized by the mixing angle $\theta_{c\bar{c}}$:
\be
\label{eta0charm}
| \eta_0 \ket = \frac{1}{\sqrt{3}} \, | u \bar{u} + d \bar{d} +
s \bar{s} \ket \, \cos \theta_{c\bar{c}} + |c \bar{c}\ket \, \sin
 \theta_{c\bar{c}} 
\quad . \ee
The orthogonal state $\eta_c$ is then given by
\be
\label{etac}
| \eta_c \ket = -\frac{1}{\sqrt{3}} \, | u \bar{u} + d \bar{d} +
s \bar{s} \ket \, \sin \theta_{c\bar{c}} + |c \bar{c}\ket \, \cos
 \theta_{c\bar{c}} 
\quad . \ee
Anticipating that the mixing angle $\theta_{c\bar{c}}$ is small, and dropping 
the $\sin^2\theta_{c\bar{c}}$ term
\footnote{If the mixing angle $\theta_{c\bar{c}}$ indeed turns out to be 
small,
the extraction of the  angle $\theta$ discussed above is not significantly
altered.},
eq.~(\ref{eta0charm}) reads approximately as 
\be
\label{eta0charmapp}
| \eta_0 \ket = \frac{1}{\sqrt{3}} \, | u \bar{u} + d \bar{d} +
s \bar{s} \ket \, \cos \theta_{c\bar{c}} + |\eta_c \ket \, \tan
 \theta_{c\bar{c}} 
\quad . \ee
$ f^{(c)}_{\eta'}$ and 
$ f_{\eta_c}$ are then related through the equation 
\be
\label{relation}
f^{(c)}_{\eta'} = \cos \theta \, \tan \theta_{c\bar{c}} \, f_{\eta_c} 
\quad , \ee
where 
$f_{\eta_c}$ is defined  
as  
$\bra \eta_c(p)|\bar{c} \g_\mu \g_5 c |0\ket = -i f_{\eta_c} \, p_\mu$.
We estimate the r.h.s. of eq.~(\ref{relation}) 
using experimental data. First, the mixing angle $\theta_{c\bar{c}}$
can be extracted
from the measured ratio
\be
\label{etac2}
R_{J/\psi}(\eta_c/\eta') \equiv
\frac{{\cal B}(J/\psi \to \eta_c \g)}{{\cal B}(J/\psi \to \eta' \g)} =
\frac{(1.3 \pm 0.4) \times 10^{-2}}{(4.31 \pm 0.30) \times 10^{-3}} \simeq
\left(\frac{k_{\eta_c}}{k_{\eta'}}\right)^3 \, 
\frac{1}{\cos^2 \theta \, \tan^2 \theta_{c\bar{c}}} \quad ,
\ee
which on using the central values of the measurements gives
$|\theta_{c\bar{c}}|=0.014$.
Second, the decay constant $f_{\eta_c}$ can be extracted
from the measured decay width {\cite{PDG96} 
\be
\label{etacdecay}
\G(\eta_c \to \g\g) = \frac{4(4\p\a)^2 \, 
f^2_{\eta_c}}{81 \p m_{\eta_c}} = 7.5^{+1.6}_{-1.4} \ \mbox{KeV}  \quad ,
\ee 
which, again for the central values, 
leads to $f_{\eta_c} = 411 $ MeV.
Eq.~(\ref{etacdecay}) is the result obtained using the standard
nonrelativistic approach. This result also follows if one assumes that the
shape of the 
distribution amplitudes for the charm and anti-charm quarks in the $\eta_c,
\eta'$ and $\eta$ mesons are very similar.  
This gives
\be
\label{fetac}
|f^{(c)}_{\eta'}| =| \cos \theta \, \tan 
\theta_{c\bar{c}} \, f_{\eta_c}| \simeq 5.8 \, \mbox{MeV} \quad.
\ee
Similarly, we can estimate the charm content
of the $\eta$ meson:
\be
\bra \eta(q)|\bar{c} \g_\mu \g_5 c |0\ket = -i f^{(c)}_{\eta} \, q_\mu
\quad ,
\ee
with 
\be
|f_\eta^{(c)}| = |\sin \theta \, \tan \theta_{c\bar{c}} \, f_{\eta_c}| 
\simeq 2.3 \, \mbox{MeV} 
\quad . 
\ee
Note that this method does not allow us to determine the signs 
of $f_{\eta'}^{(c)}$ and $f_\eta^{(c)}$,
because only the absolute value of the mixing angle
$\theta_{c\bar{c}}$ can be extracted. To illustrate this ambiguity in the 
numerical
results, we show in the case of $B^\pm \to K^\pm \eta^\prime$, the
branching ratios for both signs. 
%
%
\subsubsection{$\eta$ - $\eta'-\eta_c$ system in three-angle mixing 
formalism}

It has been recently argued that the octet-singlet mixing scheme 
involving $(\eta,\eta')$ described above needs modification 
\cite{Leutwyler97,HSLT97}. More specifically, one can show that taking into
account $1/N_c$ corrections in the effective $U(3) \otimes U(3)$ chiral
perturbation theory, one needs to distinguish the mixing angles of the octet
and singlet components. Restricting to the $(\eta,\eta')$ sector, one now
has
\be
\label{mix2}
|\eta \ket = \cos \theta_8 | \eta_8 \ket - \sin \theta_0
| \eta_0 \ket \quad ,
|\eta' \ket = \sin \theta_8 | \eta_8 \ket + \cos \theta_0
| \eta_0 \ket \quad .
\ee
The analogous relations for the pseudoscalar decay constants and masses 
can be be derived from the terms quadratic in  $\phi = 
\sum_{a=0,...,8}\lambda^a \phi_a$ in the chiral lagrangian for the 
pseudoscalar mesons $\phi_a$
(here, $\lambda^a; a=1,...,8$ are the $SU(3)$ Gell-Mann matrices and
$\lambda_0$ is a unit matrix).  
Writing
the eigenstates as $\phi_P = \sum_{a} f_P^a \phi_a$, one can show that the
vectors $(f_\eta^8,f_{\eta'}^8)$ and $(f_\eta^0,f_{\eta'}^0)$, defined 
through the matrix elements involving the axial-vector currents $A_\mu^8$ 
and $A_\mu^0$,
 \begin{eqnarray}
\langle 0 | A_\mu^8 |\eta(p) \rangle &=& i f_\eta^8 p_\mu ~, \nonumber\\
\langle 0 | A_\mu^8 |\eta'(p) \rangle &=& i f_{\eta'}^8 p_\mu ~, \nonumber\\
\langle 0 | A_\mu^0 |\eta(p) \rangle &=& i f_\eta^0 p_\mu ~, \nonumber\\
\langle 0 | A_\mu^0 |\eta'(p) \rangle &=& i f_{\eta'}^0 p_\mu ~, 
\end{eqnarray} 
are not orthogonal to each other. Instead \cite{Leutwyler97},
\be 
\label{leut1}
(f_\eta^8 f_\eta^0 + f_{\eta'}^8 f_{\eta'}^0) = -\frac{2 \sqrt{2}}{3}
(f_K^2-f_\pi^2)\left(1 + O(\delta)\right) ~,
\ee
where $O(\delta)$ represents $O(1/N_c)$ corrections.
This relation then implies
\be
\label{thetadiff}
\sin (\theta_0 - \theta_8) = \frac{2\sqrt{2}}{3} 
\frac{(f_K^2-f_\pi^2)}{f_8^2} \left(1 + O(\delta)\right) ~,
\ee
which yields (on using the decay widths $\Gamma(\eta \to 2 \gamma), 
~\Gamma(\eta' \to 2 \gamma)$ and the chiral perturbation theory result 
$f_8=1.28 f_\pi$) the following values \cite{Leutwyler97}
\be
\label{leut2}
\theta_8 = -20.5^0, ~~~\theta_0 \simeq -4.0^0 ~.
\ee
Thus, numerically the octet mixing angle $\theta_8$ comes out close to the 
angle $\theta$ but the singlet mixing angle is quite small. This implies 
that the pseudscalar meson $|\eta\rangle$ is almost a pure octet.
Extending this formalism to the $(\eta,\eta',\eta_c)$
mixing, along the lines described in the previous subsection, now yields the 
following
estimates of the quantities $f_{\eta'}^{(c)}$ and $f_{\eta}^{(c)}$:
\bea
\label{fetapc}
f_{\eta'}^{(c)} = \cos \theta_0 \tan \theta_{c\bar{c}} f_{\eta_c} \nonumber\\
f_{\eta}^{(c)} = -\sin \theta_0 \tan \theta_{c\bar{c}} f_{\eta_c} ~.
\eea

Using again the ratio 
$R_{J/\psi}(\eta_c/\eta')$ given in eq.~(\ref{etac2}) yields 
$|f_{\eta'}^{(c)}|
\simeq 5.8$ MeV (the same as before), 
but $f_{\eta}^{(c)} = - f_{\eta'}^{(c)} 
\tan \theta_0$, which is considerably smaller than the previous estimate of
the same, as $\theta_0 \ll \theta$.

 The $(\eta,\eta')$-mixing framework with two angles $\theta_0$ and $\theta_8$
fares better than the conventional one from a phenomenological point 
of view as well. Feldmann and Kroll \cite{FK97} have compared the two
mixing frameworks in a recent analysis of the $\eta \gamma$
and $\eta'\gamma$ transition form factors using data from CLEO 
\cite{CLEOetagamma}, L3 \cite{L3etagamma}, TPC \cite{TPCetagamma} and
CELLO \cite{CELLOetagamma}, the decay 
widths
$\Gamma (\eta \to 2 \gamma)$, $\Gamma (\eta' \to 2 \gamma)$ and the 
ratio $R_{J/\psi}(\eta/\eta')$. They show that even after allowing for a
moderate $SU(3)$-breaking effect, one gets a poor fit of the data in the
conventional mixing formalism (i.e., with the single angle $\theta$). 
However, the mentioned data can be well fit in the two-angle framework 
for the
$(\eta,\eta')$-mixing. Their best-fit values yield (with $f_\eta^{(c)}$ and 
$f_{\eta'}^{(c)}$ set to zero) \cite{FK97}:
 \be
\label{fkparam}
\theta_8 = -22.2^0, ~~\theta_0=-9.1^0, ~~\frac{f_8}{f_\pi} = 1.28, ~~
\frac{f_0}{f_\pi} = 1.20 ~,
\ee
which agrees reasonably well with the estimates of these parameters 
using chiral perturbation theory \cite{Leutwyler97}:
\be
\label{leut3}
\theta_8 = -20.5^0, ~~\theta_0=-4.0^0, ~~\frac{f_8}{f_\pi} = 1.28, ~~
\frac{f_0}{f_\pi} = 1.25 ~.
\ee
If accurate high-$q^2$ data were available, one could determine the
coupling constants $f_\eta^{(c)}$ and $f_{\eta'}^{(c)}$  
from the $\eta\gamma$ and 
$\eta'\gamma$ transition form factors, respectively.
 While the value
$f_\eta^{(c)} =0$ is consistent with the data, the
anaysis in \cite{FK97}  yields the following range for $f_{\eta'}^{(c)}$:
\be
\label{fkfetapc}
-65 ~\mbox{MeV} ~\leq f_{\eta'}^{(c)} ~\leq ~ 15 ~\mbox{MeV} ~.
\ee  
This
determination is somewhat model-dependent as it depends on some parameters
related to 
the charm wave-function. In the analysis reported in \cite{FK97}, the
shape of the
distribution amplitudes corresponding to the charm quark in the $\eta$ and
$\eta'$ are assumed to be the same as for $\eta_c$. It is
satisfying that the value obtained by us $|f_{\eta'}^{(c)}| \simeq 5.8$ MeV 
from $R_{J/\psi}(\eta_c/\eta')$ lies within the range given in 
eq.~(\ref{fkfetapc}).
 
 In what follows, we shall adhere to the $1/N_c$-improved chiral 
perturbation theory description of the $(\eta,\eta')$-mixing.
For numerical estimates, we  use the best-fit values given in
eq.~(\ref{fkparam}) above. We now discuss the decays $B^\pm 
(\eta,\eta')(K^\pm,K^{*\pm})$.
\subsubsection{$B^\pm \to K^\pm \eta'$}
The matrix element $M$ for $B^- \to K^- \eta'$ reads in the factorization
approximation
\bea
\label{proc6}
M &=& \frac{G_F}{\sqrt{2}} \, \left\{ V_{ub} V_{us}^* \, \left[
a_2 + a_1 \frac{m_B^2-m_{\eta'}^2}{m_B^2-m_K^2} \, 
\frac{F_0^{B \to \eta'}(m_K^2)}{F_0^{B \to K^-}(m_{\eta'}^2)} \,
\frac{f_K}{f_{\eta'}^u} \right] 
+ V_{cb}V_{cs}^* \, a_2 \frac{f_{\eta'}^{(c)}}{f_{\eta'}^u}
\right. \nonumber \\
&& - V_{tb} V_{ts}^* \, \left[2 a_3 -2 a_5 + \left( a_3 -a_5 + a_4
+ \frac{ a_6 m_{\eta'}^2}{ m_s(m_b-m_s)} \right) \,
\frac{f_{\eta'}^s}{f_{\eta'}^u} 
- \frac{ a_6 m_{\eta'}^2}{ m_s(m_b-m_s)}  \,
\right. \nonumber \\
&& \left. \left. +
\left( a_4 + \frac{2 a_6 m_K^2}{(m_s+m_u) \, (m_b-m_u)} \right)
\, \frac{m_B^2-m_{\eta'}^2}{m_B^2-m_K^2} \, \frac{F^{B \to 
\eta'}_0(m_K^2)}{F^{B \to K^-}_0(m_{\eta'}^2)} \, 
\frac{f_K}{f_{\eta'}^u} \right] \,
\right\} \, \bra K^- | \bar{s} \, b_-|B^- \ket \, 
            \bra \eta' | \bar{u} \, u_-|0 \ket \nonumber \\ 
\eea
with
\be
\label{proc6a}
 \bra K^- | \bar{s} \, b_-|B^- \ket \, 
            \bra \eta' | \bar{u} \, u_-|0 \ket =
i \, f_{\eta'}^u \, (m_B^2 - m_K^2) \, 
F_0^{B \to K^-}(m_{\eta'}^2) \quad . 
\ee 
The term proportional to $V_{cb}V_{cs}^*$ in eq.~(\ref{proc6})
is due to the charm content of the $\eta'$ as discussed above.
In eqs. (\ref{proc6}) and (\ref{proc6a})
the decay constants
$f_{\eta'}^u$ and $f_{\eta'}^s$, defined as
\be
\label{fetas}
\bra 0|\bar{u} \g_\mu \, \g_5 u |\eta' \ket =
i f_{\eta'}^u \, p_\mu \quad , \quad
\bra 0|\bar{s} \g_\mu \, \g_5 s |\eta' \ket =
i f_{\eta'}^s \, p_\mu \quad ,
\ee
are given in terms of $f_8$ and $f_0$ as
\be
\label{koppetas}
f_{\eta'}^u = \frac{f_8}{\sqrt{6}} \, \sin \theta_8 +
               \frac{f_0}{\sqrt{3}} \, \cos \theta_0  \quad , \quad
f_{\eta'}^s = -2 \, \frac{f_8}{\sqrt{6}} \, \sin \theta_8 +
               \frac{f_0}{\sqrt{3}} \, \cos \theta_0  \quad .
\ee
We remark that the matrix element $\bra 0|\bar{s} \gamma_5 s|\eta'\ket$,
which occurs when factorizing the contributions of $O_5$ and $O_6$,
has to be treated with some care. In the earlier version of the this paper
we erroneously used the relation
\be
\bra 0|\bar{s} \gamma_5 s|\eta'\ket = -i \, 
\frac{f_{\eta'}^s \, m_{\eta'}^2}{2m_s} \quad  \quad ,
\ee
which is vitiated due to the contribution of the
 anomaly term in the equation
\be
\partial^\mu  \bar{s} \gamma_\mu \gamma_5 s = 2 m_s \,
\bar{s} i \gamma_5 s + \frac{\alpha_s}{4\pi} G^{\alpha \beta}
               \tilde{G}_{\alpha \beta} \quad .
\ee
To get the correct expression for 
the matrix element $\bra \eta'|\bar{s} \gamma_5 s|0\ket$,
we now use instead the anomaly-free equation for the divergence of the
octet axial-vector current,
\be
\label{octetdiv}
\partial^\mu \left( 
\bar{u} \gamma_\mu \gamma_5 u +
\bar{d} \gamma_\mu \gamma_5 d -2
\bar{s} \gamma_\mu \gamma_5 s 
\right) = 2 i \left(
m_u \, \bar{u} \gamma_5 u +
m_d \, \bar{d} \gamma_5 d - 2 m_s \,
\bar{s} \gamma_5 s \right) \quad . 
\ee
Neglecting the terms proportional to $m_u$ and $m_d$ on the r.h.s.
of eq. (\ref{octetdiv}) one derives
\be
\bra 0|\bar{s} \gamma_5 s|\eta' \ket =
i \frac{\sqrt{6} \, f_8 \, \sin \theta_8 \, m_{\eta'}^2}{4m_s}
= -i \, 
\frac{(f_{\eta'}^s - f_{\eta'}^u ) 
\, m_{\eta'}^2}{2m_s} \quad .
\ee
Of course, this relation can also be derived by working with the
divergence of the (anomalous) singlet axial vector current. This gives
rise to the term $-a_6 m_{\eta'}^2/m_s(m_b-m_s)$ in eq.~(\ref{proc6}).
Likewise, the amplitudes of the other processes $B^- \to K^{*-} \eta'$,
$B^- \to K \eta$ and $B^- \to K^{*-} \eta$ published in the earlier 
version of this paper also get modified. The corrected amplitudes are 
given below. It appears that this (anomaly-related) error has permeated
the recent literature \cite{DDO97} and should be corrected accordingly.

  \subsubsection{$B^\pm \to K^{*\pm} 
\eta'$} The matrix element $M$ for $B^- \to K^{*-} \eta'$ 
reads in the factorization
approximation
\bea
\label{proc8}
M &=& \frac{G_F}{\sqrt{2}} \, \left\{ V_{ub} V_{us}^* \, \left[
a_2 + a_1 \frac{F_1^{B \to \eta'}(m_{K^*}^2)}{A_0^{B 
\to K^*}(m_{\eta'}^2)} \, 
\frac{f_{K^*}}{f_{\eta'}^u} \right] 
+ V_{cb}V_{cs}^* \, a_2 \, \frac{f_{\eta'}^{(c)}}{f_{\eta'}^u}
\right. \nonumber \\
&& - V_{tb} V_{ts}^* \, \left[2 a_3 -2 a_5 + \left( a_3 -a_5 + a_4
- \frac{a_6 m_{\eta'}^2}{m_s(m_b+m_s)} \right) \, 
\frac{f_{\eta'}^s}{f_{\eta'}^u} 
+ \frac{a_6 m_{\eta'}^2}{m_s(m_b+m_s)}  
\right. \nonumber \\
&& \left. \left. +
a_4
\,  \frac{F^{B \to 
\eta'}_1(m_{K^*}^2)}{A^{B \to K^*}_0(m_{\eta'}^2)} 
\, \frac{f_{K^*}}{f_{\eta'}^u} \right] \,
\right\} \, \bra K^{*-} | \bar{s} \, b_-|B^- \ket \, 
            \bra \eta' | \bar{u} \, u_-|0 \ket \nonumber \\ 
\eea 
with
\be
\label{proc8a}
\bra K^{*-} | \bar{s} \, b_-|B^- \ket \, 
\bra \eta' | \bar{u} \, u_-|0 \ket = -i \, f_{\eta'}^u \,
2 m_{K^*} \, (p_B \epsilon_{K^*}^*) \, A_0^{B \to K^*}(m_{\eta'}^2) 
\quad . 
\ee
\subsubsection{$B^\pm \to K^\pm \eta$}
The matrix element $M$ for $B^- \to K^- \eta$ reads in the factorization
approximation
\bea
\label{proc7}
M &=& \frac{G_F}{\sqrt{2}} \, \left\{ V_{ub} V_{us}^* \, \left[
a_2 + a_1 \frac{m_B^2-m_{\eta}^2}{m_B^2-m_K^2} \, 
\frac{F_0^{B \to \eta}(m_K^2)}{F_0^{B \to K^-}(m_{\eta}^2)} \,
\frac{f_K}{f_{\eta}^u} \right] 
+ V_{cb}V_{cs}^* \, a_2 \frac{f_\eta^{(c)}}{f_\eta^u} \right. \nonumber \\
&& - V_{tb} V_{ts}^* \, \left[2 a_3 -2 a_5 + \left( a_3 -a_5 + a_4
+ \frac{a_6 m_{\eta}^2}{m_s(m_b-m_s)} \right) \, 
\frac{f_{\eta}^s}{f_{\eta}^u} 
- \frac{a_6 m_{\eta}^2}{m_s(m_b-m_s)}  
\right. \nonumber \\
&& \left. \left. +
\left( a_4 + \frac{2 a_6 m_K^2}{(m_s+m_u) \, (m_b-m_u)} \right)
\, \frac{m_B^2-m_{\eta}^2}{m_B^2-m_K^2} \, 
\frac{F^{B \to 
\eta}_0(m_K^2)}{F^{B \to K^-}_0(m_{\eta}^2)} \, 
\frac{f_K}{f_{\eta}^u} \right] \,
\right\} \, \bra K^- | \bar{s} \, b_-|B^- \ket \, 
            \bra \eta | \bar{u} \, u_-|0 \ket \nonumber \\ 
\eea 
where
\be
\label{proc7a}
 \bra K^- | \bar{s} \, b_-|B^- \ket \, 
            \bra \eta | \bar{u} \, u_-|0 \ket =
i \, f_{\eta}^u \, (m_B^2 - m_K^2) \, F_0^{B \to K^-}(m_\eta^2) \quad , 
\ee 
with
\be
\label{koppeta}
f_{\eta}^u = \frac{f_8}{\sqrt{6}} \, \cos \theta_8 -
               \frac{f_0}{\sqrt{3}} \, \sin \theta_0  \quad , \quad
f_{\eta}^s = -2 \, \frac{f_8}{\sqrt{6}} \, \cos \theta_8 -
               \frac{f_0}{\sqrt{3}} \, \sin \theta_0  \quad .
\ee
\subsubsection{$B^\pm \to K^{*\pm} \eta$}
The matrix element $M$ for $B^- \to K^{*-} \eta$ 
reads in the factorization
approximation
\bea
\label{proc9}
M &=& \frac{G_F}{\sqrt{2}} \, \left\{ V_{ub} V_{us}^* \, \left[
a_2 + a_1 
\frac{F_1^{B \to \eta}(m_{K^*}^2)}{A_0^{B \to K^*}(m_{\eta}^2)} \, 
\frac{f_{K^*}}{f_{\eta}^u} \right] 
+ V_{cb}V_{cs}^* \, a_2 \frac{f_\eta^{(c)}}{f_\eta^u}
\right. \nonumber \\
&& - V_{tb} V_{ts}^* \, \left[2 a_3 -2 a_5 + \left( a_3 -a_5 + a_4
- \frac{a_6 m_{\eta}^2}{m_s(m_b+m_s)} \right) \, 
\frac{f_{\eta}^s}{f_{\eta}^u} 
+ \frac{a_6 m_{\eta}^2}{m_s(m_b+m_s)}  
\right. \nonumber \\
&& \left. \left. +
a_4
\,  \frac{F^{B \to 
\eta}_1(m_{K^*}^2)}{A^{B \to K^*}_0(m_{\eta}^2)} 
\, \frac{f_{K^*}}{f_{\eta}^u} \right] \,
\right\} \, \bra K^{*-} | \bar{s} \, b_-|B^- \ket \, 
            \bra \eta | \bar{u} \, u_-|0 \ket \nonumber \\ 
\eea 
with
\be
\label{proc9a}
\bra K^{*-} | \bar{s} \, b_-|B^- \ket \, 
\bra \eta | \bar{u} \, u_-|0 \ket = -i \, f_{\eta}^u \,
2 m_{K^*} \, (p_B \epsilon_{K^*}^*) \, A_0^{B \to K^*}(m_\eta^2) \quad . 
\ee
\section{Input Parameters, Numerical Results and Comparison with the CLEO 
Data}
 \setcounter{equation}{0}
\subsection{Input parameters}
  The matrix elements for the decay $B \to h_1 h_2$ derived 
in the preceding section depend on the
effective coefficients $a_1,...,a_6$, quark masses, 
various form factors, coupling constants and the CKM parameters.
In turn, 
the  coefficients $a_i$ and the quark masses depend on 
the renormalization scale $\mu$ and
the QCD scale parameter $\Lambda_{\overline{\mbox{MS}}}$.
We have
fixed $\Lambda_{\overline{\mbox{MS}}}$ using $\alpha_S (M_Z)=0.118$, which is
the central value of the present world average
 $\alpha_S (M_Z)=0.118 \pm 0.003$ \cite{Schmelling96}. The 
scale $\mu$ is varied
between $\mu =m_b$ and $\mu =m_b/2$, but due to the inclusion of the NLL
expressions, the dependence of the decay rates on $\mu$ is small and hence 
not pursued any further. 
To be specific, we use $\mu=2.5$ GeV in the following.
The CKM matrix will be
expressed in terms of the Wolfenstein parameters \cite{Wolfenstein83}, 
$A$, $\lambda$, $\rho$ and the phase $\eta$. Since the first two are 
well-determined
with $A= 0.81 \pm 0.06, ~\lambda=\sin \theta_C=0.2205 \pm 0.0018$, we fix
them to their central values. The other two are correlated and are found 
to lie (at 95\% C.L.) in the range $0.25 \leq \eta \leq 0.52$ and
$-0.25 \leq \rho \leq 0.35$ from the CKM unitarity fits
\cite{AL96}. We take four representative points in the
allowed $(\rho,\eta)$ contour. Their values and the legends used in drawing
the figures  are as follows:
\begin{enumerate}
\item $\rho = 0.05, \eta = 0.36$, yielding $\sqrt{\rho^2 + \eta^2} =0.36 $
(drawn as a solid curve)
\item $\rho = 0.30, \eta = 0.42$, yielding 
$\sqrt{\rho^2 + \eta^2} =0.51 $ (drawn as a dashed curve)
\item $\rho = 0, \eta = 0.22$, yielding
$\sqrt{\rho^2 + \eta^2} =0.22 $ (drawn as a dashed-dotted curve)
\item $\rho = -0.20, \eta = 0.45$, yielding
$\sqrt{\rho^2 + \eta^2} =0.49 $ (drawn as a dotted curve).
\end{enumerate}
They correspond to the central values of the fits in \cite{AL96}, maximum
 allowed value of $|V_{ub}/V_{cb}|$ with positive 
$\rho$, minimum allowed value of $|V_{ub}/V_{cb}|$, and maximum allowed
value of $|V_{ub}/V_{cb}|$ with negative $\rho$, respectively.
The CKM parameters are also an output from the measured
non-leptonic $B$ decays and we shall illustrate the potential
interest in this kind of analysis using some of the ratios of
the branching ratios as an exercise. The rest of the input quantities
used in our estimates for the branching ratios are collected
in several tables. We discuss now these input values.

\subsubsection{Effective coefficients in the factorization scheme}

With the electroweak penguins and the so-called $W$-annihilation/exchange
diagrams neglected, the amplitudes for the various decays depend on
 six coefficients, $a_i$, defined in 
section 3. Eventually, one should determine each one of them
(or particular combinations thereof) by analyzing the specific 
decay modes most sensitive to these coefficients. This way, one can
measure the deviation in each one of them from their values in perturbation 
theory and determine if this deviation (due to non-perturbative 
effects) can be described in terms 
of a few universal parameters. Perhaps, it should be remarked here that an
analysis of the heavy to heavy transitions in two-body $B$ decays can be
reasonably well described in terms of one parameter, called $\zeta$ in 
\cite{NS97}, whose value seems to be universal.
Following this, we do the simplest thing here by assuming
that a single parameter $\xi$, defined in the preceding section, can be 
used  to compensate for neglecting the octet-octet terms in all
matrix elements of the decays $B \to h_1 h_2$. This is motivated by the
fact that the energy release in these decays is  comparable.
It remains an open question if the parameter $\xi$ introduced
here in the decays such as $B \to K \pi$ is close to the corresponding
parameter $\zeta$, entering, for example, in the decay $B \to D \pi$.  
We show the  dependence of the branching ratios in several decay modes in 
the range $0 \leq \xi \leq 1$, with $\xi=1/3$ being the naive factorization
value, i.e., if one uses factorization and neglects the octet-octet 
contribution in the matrix elements.

\subsubsection{Decay coupling constants and form factors}
 
For the various decay constants occurring in the formulas in section
3 we use the numerical values shown in table 2.
\begin{table}[htb]
\label{decayconst}
\begin{center}
\begin{tabular}{| r | r | r | r | r | r | r | r|}
\hline
 $f_\omega$ & $f_K$  & $f_{K^*}$ & $f_\pi$ & $f_0$ & $f_8$ 
& $|f_{\eta'}^{(c)}|$ &  $|f_{\eta}^{(c)}|$\\
 \hline
195 & 160 & 214 & 131 & 157  & 168 &
5.8 & 0.93\\
\hline
\end{tabular}
\end{center}
\caption{Decay constants in MeV.}
\label{decayconsttab}
\end{table}
The values for $f_\omega, f_K, f_{K^\ast}$ and $ f_\pi$ coincide with the 
central values quoted in \cite{NS97} extracted from data on the
electromagnetic decays of $\omega$ and 
$\tau$ decays, respectively \cite{PDG96}.
The decay constants $f_{\eta'}^u$,  $f_{\eta'}^s$, 
$f_{\eta}^u$ and $f_{\eta}^s$ defined in eqs. 
(\ref{koppetas}) and (\ref{koppeta}) are obtained from the 
values for $f_0$ and $f_8$ in table 2 and by using 
$\theta_8=-22.2^o$ for the $(\eta,\eta')$ mixing angle \cite{GK87}. The errors
on the coupling constants in the table are small 
(typically $(1 - 3)\%$), except
on $f_{\eta'}^{(c)}$ and $f_{\eta}^{(c)}$, for which present data allow a 
determination with an error of $\pm 15\%$ (assuming the mixing formalism
holds). 

The decays being considered here, such as $B \to \pi K$, involve light
hadrons in the final state. The rates require the knowledge of the various
form factors at  $q^2=m_h^2$, where $m_h$ denotes a light hadron mass.
Since $q^2=m_h^2$ is rather close to the point $q^2=0$, and a simple pole 
model is mostly used to implement the $q^2$ dependence in the form factors,
we shall neglect this $q^2$-dependence in the form factors and equate 
$F_{0,1}^{B \to h} (q^2 =m_h^2) =
F_{0,1}^{B \to h} (q^2 =0)$. Explicit calculations bear this out 
and find that the variation in the stated 
range is indeed  small \cite{NS97,ABS94,BB97}. The values used for 
the form factors $F_{0,1}^{B \to h}(q^2=0)$ 
and $A_0^{B \to h}(q^2=0)$ in our rate estimates are listed in 
Table 3. They are taken from \cite{BSW87}, which are reproduced in most
other calculations (see, for example, Table 1 in \cite{BB97}). Note also,
that the $SU(3)$-breaking effects in the form factors are neglected. They
are typically of $O(20)\%$ \cite{ABS94}.
\begin{table}[htb]
\label{formfac}
\begin{center}
\begin{tabular}{| r | r | r | r | r | r | r |}
\hline
 $F_{0,1}^{B \to K^-}$ & $F_{0,1}^{B \to \pi^-}$  
& $F_{0,1}^{B \to \pi^0}$ & $F_{0,1}^{B \to \eta'}$ 
& $F_{0,1}^{B \to \eta}$ 
& $A_0^{B \to \omega}$  & $A_0^{B \to K^*}$ \\
 \hline
0.33 & 0.33 & $\frac{0.33}{\sqrt{2}}$ & 
$0.33 \left[ \frac{\sin \theta_8}{\sqrt{6}} + \frac{\cos \theta_0}{\sqrt{3}} 
\right] $ &
$0.33 \left[ \frac{\cos \theta_8}{\sqrt{6}} - \frac{\sin \theta_0}{\sqrt{3}} 
\right] $
& $\frac{0.28}{\sqrt{2}}$ & 0.28 \\
\hline
\end{tabular}
\end{center}
\caption{Form factors at $q^2=0$.}
\label{table3}
\end{table}

\subsubsection{Current and constituent quark masses}
The quark masses enter our analysis in two different ways. 
First, they occur in the amplitudes involving penguin loops.  
We treat the internal quark masses in these loops
as constituent masses rather than current masses. For them we use the
following (renormalization scale independent) values:
\begin{equation}
\label{constmasses}
m_b=4.88 ~\mbox{GeV}, ~~m_c=1.5 ~\mbox{GeV}, ~~m_s=0.5 ~\mbox{GeV},  
~m_d=m_u=0.2  ~\mbox{GeV}.
\end{equation}
Variation in a reasonable range of these parameters does not change
the numerical results of the branching ratios significantly. The value of 
$m_b$ above is fixed to be
the current quark mass value $\overline{m_b}(\mu=m_b/2)=4.88$ GeV, given
below.
 Second, the quark masses $m_b$, $m_s$, $m_d$ and $m_u$  
also appear through the equations of motion
when working out the (factorized) hadronic matrix elements.
In this case, the quark masses should be interpreted as current masses.
Using $\overline{m_b}(m_b)=4.45 \, $ GeV \cite{GKL96}
and 
\be
\label{msbarmass}
\overline{m_s}(1 \ GeV) = 150 \ MeV \quad , \quad 
\overline{m_d}(1 \ GeV) = 9.3 \ MeV \quad , \quad 
\overline{m_u}(1 \ GeV) = 5.1 \ MeV \quad , 
\ee
from \cite{GaL85}, the corresponding values at the renormalization
scale $\mu=2.5$ GeV are given in table 4, together with other
input parameters needed for our analysis.
\begin{table}[htb]
\label{formfact}
\begin{center}
\begin{tabular}{| r | r | r | r | r | r | r | r |r|}
\hline
 $\overline{m_b}$ & $\overline{m_s}$  & 
$\overline{m_d}$ & $\overline{m_u}$ & $\a_s(m_Z)$ 
& $\tau_B$ & $\bra q^2 \ket$ & $\theta_8$ & $\theta_0$ \\
 \hline
4.88 GeV & 122 MeV & 7.6 MeV & 4.2 MeV & 0.118 & 1.60 ps & $m_b^2/2$ &
$-22.2^o$ & $-9.1^0$\\
\hline
\end{tabular}
\end{center}
\caption{Quark masses and other input parameters. The running masses
are given at the renormalization scale $\mu=2.5$ GeV.}
\label{table4}
\end{table}
\subsubsection{Numerical values for the effective Wilson coefficients 
$C_i^{eff}$}
From eqs.~(\ref{ct}) - (\ref{cg}) it follows that the effective Wilson
coefficients $C_i^{eff}$ defined in eq.~(\ref{ceff}) 
are in general complex numbers, which
depend on quarks masses and on the CKM matrix elements.
Taking the quark masses listed in eq. (\ref{constmasses})
and using the central values for the CKM parameters from the unitarity fits
\cite{AL96}
(i.e. $\rho=0.05$ and $\eta=0.36$), the effective Wilson coefficients
$C_i^{eff}$ at the renormalization scale $\mu=2.5$ GeV are shown in table
5. We remark that the (almost) identical values of these coefficients
in the first two columns ($b \to s$ and $\bar{b} \to \bar{s}$) reflects that 
the imaginary parts of these 
effective Wilson coefficients are essentially generated by strong 
interactions.
The numerically differing entries in the other two columns ($b \to d$ and 
$\bar{b} \to \bar{d}$) 
reflect that the weak (CP-violating) and strong interaction phases
in these decays are comparable. 
  \begin{table}[htb]
\begin{center}
\begin{tabular}{| r | c | c | c | c |}
\hline
           & $b \to s$  & $\bar{b} \to \bar{s}$ &
             $b \to d$  & $\bar{b} \to \bar{d}$ \\ 
 \hline
$C_1^{eff}$ & $1.160$   & $1.160$    & $1.160$     & $1.160$   \\
$C_2^{eff}$ & $-0.334$ & $-0.334$ & $-0.334$ & $-0.334$ \\
$C_3^{eff}$ & $0.021+0.004i$& $0.021+0.004i$& $0.020+0.002i$& $0.022+0.006i$ \\
$C_4^{eff}$ & $-0.052-0.011i$ &$-0.051-0.011i$&$-0.048-0.007i$&$-0.053-0.017i$ 
\\
$C_5^{eff}$ & $0.016+0.004i$ &$0.016+0.004i$&$0.015+0.002i$&$0.017+0.006i$ \\
$C_6^{eff}$ & $-0.064-0.011i$&$-0.063-0.011i$&$-0.060-0.007i$&$-0.065-0.017i$\\
\hline
\end{tabular}
\end{center}
\caption{Effective Wilson coefficients $C_i^{eff}$
at the renormalization scale $\mu=2.5$ GeV for the various
$b \to q$ ($\bar{b} \to \bar{q}$) transitions. See text and eq. (\ref{ct}).}
\label{table5}
\end{table}


\subsection{Numerical results and comparison with CLEO data}
 Having stated our theoretical framework and the input parameters, we
now present our results for the various decays of interest listed in the
previous section. A word of caution concerning the accuracy 
of the absolute decay rates calculated by us is in order. As just 
displayed, there are 
many parameters involved in describing exclusive non-leptonic decays and
while the decay rates do not depend sensitively on all of them,
and many input parameters are already well known, it is
obvious that the predicted branching ratios 
do depend sensitively on some for which there is no alternative at present
to using model-dependent estimates.
 The particular quantities in question are
the decay form factors.
Some of these form factors enter in other processes
which have been measured (such as in the semileptonic and radiative $B$ 
decays) and the estimates being used are found to reproduce the data quite 
well, yet some others are not yet constrained by data directly. So, the
estimates given below 
for the absolute decay rates have to be taken with an accuracy which is not 
better than a factor 2.
The additional uncertainty due to the parameter $\xi$ can not be judged
at this stage. That can only be ascertained in future, if this framework
proves to be a reasonable way to analyze heavy to light transitions in 
$B$ decays.

 However, within this framework, the ratios of the branching ratios are much
more stable, as many of the theoretical uncertainties (such as in the form
factors, various scales, and quark masses) cancel out to a large extent.
In some cases, the dependence on the parameter $\xi$ also cancels, 
or it is
very weak. Hence, the ratios are
more reliable and the experimental information on these ratios can 
eventually be
used meaningfully to draw inferences on the fundamental parameters, such
as $\rho$ and $\eta$.

\subsubsection{Branching ratios for $B \to \pi \pi$ modes}
We shall show the branching ratios of interest as a function of the
parameter $\xi$ for four different set of values of the CKM parameters.
Wherever available, the present measurements of the branching ratios at the
$\pm 1 \sigma$ level are also shown on these figures (thick solid 
lines).  
All experimental numbers are taken from
\cite{Behrens97,Jsmith97,Wuerthwein97}, and in showing the experimental
results, we have added the statistical and systematic errors in quadrature.
 We start by showing in Fig.~\ref{Bpmpipi} the branching ratio    
${\cal B}(B^\pm \to \pi^0 \pi^\pm)$. The decay rate for this mode is
sensitive to both the variation in $\xi$
and  the CKM parameters. This is obvious from the quadratic dependence
of the decay rate on the quantity $|V_{ub}|$. Also, it 
depends on the combination $a_1 + a_2$. Hence, a measurement of this decay
rate will yield information on these quantities. In quoting a range, we
shall take $0 \leq \xi \leq 0.5$ (which is suggested by the combined 
analysis of all the present CLEO data on $B \to h_1 h_2$ decays which we
show later). We estimate, 
$${\cal B}(B^\pm \to \pi^0 \pi^\pm) \simeq (0.1 - 1.4) \times 10^{-5},$$
which is uncertain by over an order of magnitude. However, the lower
range corresponds to the rather small value of the CKM-factor,  
$|V_{ub}/V_{cb}|= 0.05$,
and is therefore somewhat unlikely.  For the central value 
$|V_{ub}/V_{cb}|= 0.08$, we estimate 
$${\cal B}(B^\pm \to \pi^0 \pi^\pm) =(0.3 - 0.6) \times 10^{-5}.$$
The present experimental upper limit is (at $90\% $ C.L.)
 $${\cal B}(B^\pm \to \pi^0 \pi^\pm) < 2.0 \times 10^{-5}.$$

 In Fig.~\ref{Bzpippim}, we show the branching ratio 
${\cal B}(B^0 (\overline{B^0}) \to 
\pi^\mp \pi^\pm)$. Again, this decay mode is sensitive to $\xi$ and
the CKM parameters, although the resulting uncertainty is less in this 
case than in ${\cal B}(B^\pm \to \pi^0 \pi^\pm)$.
 Comparison of the model calculations with the
present upper limit (at $90\% $ C.L.)
 $${\cal B}(B^0 (\overline{B^0}) \to \pi^\mp \pi^\pm) < 1.5 \times 10^{-5}.$$
shows that this decay mode is expected to lie within a factor $2-3$
 of the 
present upper limit and hence should be measured soon. 
Already, the present upper limit on this mode disfavors some extreme 
values of the CKM parameters corresponding to $|V_{ub}/V_{cb}|$
close to or in excess of $0.11$.

 In Fig.~\ref{Bzpi0pi0}, we show the branching ratio ${\cal B}
(B^0 (\overline{B^0})\to
\pi^0 \pi^0)$. This branching ratio is not very sensitive to $\xi$ in
the region $0 \leq \xi \leq 0.5$ but rises sharply as $\xi \to 1$. All
the curves lie however significantly below the present upper limit
\cite{PDG96}:
 $${\cal B}(B^0 (\overline{B^0}) 
\to \pi^0 \pi^0) < 9.1 \times 10^{-6}.$$
Restricting to $0 \leq \xi \leq 0.5$, our model calculation yields
$${\cal B}(B^0 
(\overline{B^0}) \to \pi^0 \pi^0) \simeq  (0.5 - 2.0) \times 10^{-6}.$$

\subsubsection{Branching ratios for $B \to \pi K$ modes}

   In Fig.~\ref{Bpmpik}, we show the branching ratio
${\cal B}(B^\pm \to \pi^\pm K)$. This is a good decay mode, in principle, to 
determine
the parameter $\xi$, as there is no perceptible dependence of the rate
on the CKM parameters. In the indicated range $0 \leq \xi \leq 1$, the
branching ratio varies by slightly more than a factor 2.
The experimental measurement is (at $\pm 3.2\sigma$):
$${\cal B}(B^\pm \to \pi^\pm K) = (2.3 ^{+1.1 + 0.2}_{-1.0 -0.2} \pm
0.2) \times 10^{-5}.$$
Our estimated branching ratio is in agreement with data, and there is a
slight preference for smaller values of  $\xi$, with  $\xi >0.7 $
somewhat disfavored. Since the CKM-parametric dependence is small,
this decay mode is useful to show the effects of the QCD corrections.
 In Fig.~\ref{Bkpinll}, we show the branching ratio ${\cal B}(B^\pm 
\to \pi^\pm K)$ as a function of the scaled variable  $\langle 
q^2/m_b^2\rangle$, in the range $0 \leq \langle q^2/m_b^2\rangle \leq 1$,
calculated for $\xi =0$.
The dashed line corresponds to the LL approximation, whereas the dotted 
and solid lines correspond to the truncated NLL approximation, 
and the complete NLL approximation as discussed in section 2, respectively.
The dotted curve amounts to what has been used in the analysis of
the decay modes $(B^\pm \to \pi^\pm K)$ in \cite{FM97-1, FM97-2}. The effect 
of the
complete NLL corrections is numerically important, and they tend to decrease
the branching ratio as compared to what one estimates by including the
charm penguins alone.

   In Fig.~\ref{Bzpik}, we show the branching ratio
${\cal B}(B^0 (\overline{B^0}) \to \pi^\pm K^\mp)$. Like its charged partner,
${\cal B}(B^\pm\to \pi^\pm K)$ discussed above, 
this decay mode is also sensitive to
the parameter $\xi$, though in this case there is a perceptible dependence
of the rate on the CKM parameters as well.
 The observed branching 
ratio (at $\pm 5.6\sigma$):
$${\cal B}(B^0 (\overline{B^0}) \to \pi^\pm K^\mp) =
 (1.5 ^{+0.5 +0.1}_{-0.4- 0.1} 
\pm 0.1) \times 10^{-5},$$
is quite comfortably accommodated by our estimates. 

Comparing Figs.~\ref{Bzpik}
and ~\ref{Bpmpik}, one sees that the dependence of these decay rates on
$\xi$ is very similar, and hence in the ratio of branching ratios it
almost cancels out. Defining this ratio by $R_1$,
\begin{equation}
R_1 \equiv \frac{{\cal B}(B^0 (\overline{B^0})
\to \pi^\pm K^\mp)}{{\cal B}(B^\pm \to 
\pi^\pm K)},
\end{equation}
we show $R_1$ as a function of the CKM parameter $\rho$ in
 Fig.~\ref{R1} for two values of the CKM parameter $\eta=0.52$ (upper
curve) and $\eta=0.25$ (lower curve).  We note that $R_1$ is rather
 insensitive to $\eta$ but it
does depend sensitively on $\rho$. Using the
present CLEO measurement of $R_1$ (at $\pm 1 \sigma$)
\begin{equation}
R_1= 0.65 \pm 0.40 \quad ,  
\end{equation}
Fig.~\ref{R1} suggests that negative values of $\rho$ are disfavored. 
This can also be converted as a statement 
on the CP-violating phase $\gamma$. Since the Wolfenstein parameter $\eta$
is positive as determined from the constraint on $\epsilon_K$, 
$\rho > 0$ implies $\gamma < 90^\circ$.
 We recall that the  bounds on $\gamma$ obtained from
the CKM unitarity fits yield symmetric constraints, centred around
$\gamma=90^0$ (or $\rho=0$). However, it should be
remarked that the lower bound on the ratio of the weak mass differences
in the $B_s^0$-$\overline{B_s^0}$ and $B_d^0$-$\overline{B_d^0}$ systems, 
$\Delta M_s/\Delta M_d$, which at present is posted as
$\Delta M_s/\Delta M_d > 20.4$ at $95\%$ C.L. \cite{HGMoser97}, now cuts
away a good part of the negative-$\rho$ (equivalently $\gamma > 90^o$) 
region. A recent analysis 
gives (at $95\%$ C.L.): $32^0 \leq \gamma \leq 122^0$ \cite{AL97}, which is
no longer symmetric around $\gamma=90^0$. 
 On the other hand, the 
model-independent constraints on $\gamma$ from $R_1$, 
discussed by Fleischer and Mannel \cite{FM97-1},
are such that they force $\gamma$ to lie in the range $0^0 \leq \gamma \leq
\gamma^{max}$ or $180^0-\gamma^{max} \leq \gamma \leq 180^0$, depending on
the sign of $\cos \delta$, where $\delta$ is the strong phase-shift 
difference between the tree and penguin amplitudes in the decay
$B^0(\overline{B^0}) \to \pi^\pm K^\mp$ . Since
this phase difference is calculated in our model, 
the preferred solution is 
the one in which $\gamma$ lies in the first quadrant, or $0^0 \leq \gamma 
\leq \gamma^{max}$. Unfortunately, with the
present experimental errors, the $95\%$ C.L. limit on $\rho$ from $R_1$
(or on $\gamma^{max}$) does not allow one to draw more quantitative
conclusions on the value of $\gamma$  than
what one gets from the CKM fits \cite{AL96,AL97}. This is
expected to change with improved data on $R_1$, if the value of $R_1$ is
found to be considerably less than 1.
Our analysis,
carried out in the factorization framework,  
underlines the sensitive dependence of $R_1$ on $\rho$, with $\rho \leq 0$ 
disfavored (at $\pm 1\sigma$) by the CLEO data on $R_1$. The 
effect of the present lower bound on $\Delta M_s/\Delta M_d$ on $\rho$ is 
qualitatively similar to the one
from the present measurement of $R_1$, namely both prefer $\rho \geq 0$.
For an updated CKM fits, see also \cite{PPRS97}.
 
\subsubsection{Branching ratios for the $B \to h^\pm \pi$ and $B \to 
h^\pm K$ modes}
 The decay modes $B^\pm \to h^\pm \pi^0$, 
$B^0 (\overline{B^0})  \to h^\pm \pi^\mp$
and  $B^\pm \to h^\pm K^0$ have been measured with impressive precisions.
We compare our model estimates with these measurements.
   In Fig.~\ref{Bpmpihpm}, we show the branching ratio
${\cal B}(B^\pm \to  h^\pm\pi^0)$. The decay rate in this case is mildly
dependent on $\xi$, but more importantly on the CKM parameters. 
 The experimental measurement (at $\pm 5.5\sigma$):
$${\cal B}(B^\pm \to \pi^0 h^\pm ) = (1.6 ^{+0.6+ 0.3}_{-0.5 -0.2} \pm
0.1) \times 10^{-5}$$
 is reproduced well by our model.

 In Fig.~\ref{Bzhpmpi}, we compare our model estimates with the CLEO 
measurements (at $\pm 7.8\sigma$):
$${\cal B}(B^0 (\overline{B^0}) \to h^\pm \pi^\mp ) = 
(2.2 ^{+0.6}_{-0.5} \pm 0.1) \times 10^{-5}.$$
Agreement between our model and data is good. The two curves (dashed and 
dotted) which lie outside the $\pm 1 \sigma$ bands correspond to large
values of the ratio $|V_{ub}/V_{cb}|$, namely $|V_{ub}/V_{cb}| =0.11$, which 
is also outside of the $\pm 1 \sigma$ bound from direct measurements of
$|V_{ub}/V_{cb}|$. So, all of these different pieces of data are giving a
consistent picture.  

In Fig.~\ref{Bpmhpmk}, we show our estimates for the branching ratio
for the mode $ B^\pm \to h^\pm K^0$, which has been measured (at $4.4 \sigma$) 
$${\cal B}(B^\pm \to h^\pm K^0 ) = (2.4 ^{+1.1 +0.2}_{-1.0 - 0.2} \pm
0.2) \times 10^{-5}.$$ This branching ratio has a very similar dependence on
$\xi$ as in the decay $B^\pm \to \pi^\pm K^0$ and likewise has little
dependence on the CKM parameters. Model estimates are in agreement with 
data for $\xi \leq 0.7$.  

As another example of a ratio of branching ratios, which is sensitive to
the CKM parameters, we define the ratio $R_2$
\begin{equation}
R_2 \equiv \frac{{\cal B}(B^0 (\overline{B^0}) 
\to h^\pm \pi^\mp)}{{\cal B}(B^\pm \to
\pi^\pm K^0)},
\end{equation}
which like $R_1$ is less dependent on the
other input parameters, including $\xi$. Since ${\cal B}(B^\pm \to  
\pi^\pm K^0)$ is insensitive to the CKM parameters, the ratio $R_2$ 
reflects 
the CKM
dependence of  ${\cal B}(B^0 (\overline{B^0}) \to h^\pm \pi^\mp)$. We plot
the ratio $R_2$ in Fig.~\ref{R2} as a function of the phase $\eta$, for
three values of $\rho$: $\rho =0.05$ (dashed curve), $\rho = 0.35$ (solid
curve) and $\rho =-0.25$, which coincides with the case $\rho=0.35$.
The present experimental value of $R_2$ (at $\pm 1 \sigma$) is
\begin{equation}
R_2= 0.96 \pm 0.57 \quad .
\end{equation}
This shows that with the stated significance $R_2$ disfavors large 
values of $\eta$ in excess of $\eta \geq 0.5$.

\subsubsection{Branching ratios for the $B^\pm \to \omega K^\pm$ and $B^\pm 
\to \omega h^\pm$ modes}

 Next, we study the decays $B^\pm \to \omega K^\pm$ and $B^\pm \to \omega 
h^\pm$ ($h=\pi,K$), which have also been measured by the CLEO collaboration
\cite{Behrens97}, with the former having a branching ratios (at $\pm 3.3 
\sigma)$ 
 $${\cal B}(B^\pm \to \omega K^\pm ) = (1.2 ^{+0.7}_{-0.5} \pm
0.2) \times 10^{-5},$$
and the latter (at $\pm 6.0 \sigma)$
 $${\cal B}(B^\pm \to \omega h^\pm ) = (2.5 ^{+0.8}_{-0.7} \pm
0.5) \times 10^{-5}.$$
These measurements are compared with our model calculations in
Figs.~\ref{Bpmomegakpm} and \ref{Bpmomegahpm}, respectively. Both of these
decays have  an interesting dependence on 
the variable $\xi$. Taken the data at face value ($\pm 1 \sigma)$, a
value for $\xi$ in the range $0.15 \leq \xi \leq 0.5$ and $\xi \geq 
0.85$ are somewhat disfavored by data in the decay $B^\pm \to \omega K^\pm$.
 Curiously, the estimated branching ratio
${\cal B}(B^\pm \to \omega K^\pm )$ has its lowest value in the range
$\xi = 0.3 \pm 0.1$, and in this range it fails to reproduce the 
data by almost $2 \sigma$. 
 This observation and the present measurement of
${\cal B}(B^\pm \to \pi^\pm K)$ as well as ${\cal B}(B^\pm \to h^\pm K)$,
which disfavor $\xi \geq 0.7$ 
then imply that the preferred value of
$\xi$ in our model is either in the range $0 \le \xi \le 0.15$, or else
$\xi \simeq 0.5$. In this range, however,
the estimated branching ratio is somewhat lower than the experimental one
in $B^\pm \to \omega h^\pm$, but not by a large amount.
Due to the fact that the data being discussed are the first ones
of their kind and the uncertainties related to the parameters of the
present theoretical framework have not been exhaustively studied, one can 
not draw too strong conclusions on the value of the parameter $\xi$ from
this decay. 

\subsubsection{Branching ratios for the $B^\pm \to (\eta,\eta^\prime) 
(K^\pm, K^{\ast \pm})$ modes}
Finally, we take up the decay $B^\pm \to \eta^\prime K^\pm$, which has 
attracted a lot of theoretical attention recently. Compared to the decays 
considered so far, this decay and the related ones $B^\pm \to \eta 
K^\pm$, $B^\pm \to \eta^\prime K^{\ast \pm}$ and $B^\pm \to \eta K^{\ast 
\pm}$ have  an extra contribution  from  the decay
chain $b \to s c \bar{c} \to  s(\eta, \eta^\prime)$.

In Fig.~\ref{Bpmetapkpm} we show the branching ratio ${\cal B}(B^\pm 
\to \eta^\prime K^\pm)$ as a function of $\xi$, varying the CKM 
parameters as 
indicated in section 4.1. Since we are not able to 
determine the sign of the coupling constant $f_{\eta^\prime}^{(c)}$ due to
the sign ambiguity in the determination of the angle $\theta_c$, 
we show the
result for both $f_{\eta^\prime}^{(c)} = \pm 5.8$ MeV. Note that the 
$\xi$-dependence of this branching ratio results in a factor 2 
uncertainty varying $\xi$ in the 
range $0 \leq \xi \leq 0.5$ for the positive-$f_{\eta'}^c$ solution; the
branching ratio is less sensitive to $\xi$ for the negative-$f_{\eta'}^c$ 
case.
The positive-$f_{\eta^\prime}^{(c)}$ solution yields a marginally higher
branching ratio.
The CKM-parametric dependence of this branching ratio is not very marked.
 Within the present uncertainties in the 
input parameters, we get, at $\xi\simeq 0$ 
$${\cal B}(B^\pm \to \eta^\prime K^\pm) \simeq (3 - 4) \times 10^{-5},$$
which at $\xi =0.5$ falls down to the range 
$${\cal B}(B^\pm \to \eta^\prime K^\pm) \simeq (2 - 3) \times 10^{-5}.$$
This is  
to be compared with the CLEO measurement (at $\pm 5.5 \sigma)$
$${\cal B}(B^\pm \to \eta^\prime K^\pm) = (7.1 ^{+2.5}_{-2.1} \pm 0.9) 
\times 10^{-5}.$$
Given the experimental and theoretical errors, the model estimates and data 
are clearly not incompatible though, with the values of the parameters 
used by us, our estimates are somewhat on the
lower side. Since, apart from the form factors, this branching ratio is 
sensitive to the value of the
$s$-quark mass, with the branching ratio increasing as $m_s$ is decreased 
from its default value $m_s(\mu =2.5 ~\mbox{GeV})=122$ MeV used by us, 
the decay rate can be made to agree with the CLEO data by optimising
these parameters in an overall fit. This is not warranted at this stage. 

The branching ratios for the decays  $B^\pm \to \eta^\prime K^{\ast \pm}$,
 $B^\pm \to \eta K^\pm$ and  $B^\pm \to \eta K^{\ast \pm}$ are 
shown in Figs.~\ref{Bpmetapkstpma}, 
 \ref{Bpmetakpm}, and \ref{Bpmetakstpm}, respectively, for the values
$f_{\eta^\prime}^{(c)}=-5.8$ MeV and 
$f_{\eta}^{(c)}=-0.93$ MeV.
The reason for selecting the negative sign  is to be traced to the
observation that the contribution of the amplitude $b \to (c\bar{c}) s
\to (\eta,\eta') s$ can also be calculated using QCD-anomaly which fixes
the signs of these constants to be negative \cite{ACGK97}.  
The estimated branching ratios satisfy the 
respective present experimental
bounds on them \cite{Behrens97}.
For $0 \leq \xi \leq 0.5$, we predict
$${\cal B}(B^\pm \to \eta^\prime K^{\ast\pm}) \simeq (0.3 - 0.9) \times 
10^{-6}.$$
The decays $B^\pm \to \eta (K^\pm,K^{\pm *})$
on the other hand do not depend very sensitively on the sign
of $f_{\eta}^{(c)}$. We estimate ($0 \leq \xi \leq 0.5$) 
$${\cal B}(B^\pm \to \eta K^{\pm}) \simeq (1.0 - 2.8) \times 10^{-6},$$
$${\cal B}(B^\pm \to \eta K^{\ast \pm}) \simeq (1.0 - 2.8) \times 10^{-6}.$$

 Finally, we remark that scenarios
with a greatly enhanced strength of the dipole operator $O_8$ have
been entertained in the literature \cite{Kagan95,CGG96,LNO97}, with the
view of bringing the existing
theoretical estimates of the semileptonic branching ratios and charm
counting in $B$ decays in better rapport with data. A
greatly enhanced value of $C_8(m_W)$ will influence the
branching ratios in some selected non-leptonic $B$ decay channel as 
well.   
However, this effect is diluted due to the contributions from other
Wilson coefficients, which are assumed to have their SM values. Also, as  
emphasized in \cite{Kagan95,CGG96}, and more recently in \cite{LNO97}, the
strong mixing of the operators $O_2$ and $O_8$ would require a very   
large enhancement in $C_8(m_W)/C_{8}^{SM}(m_W)$, typically
$O(10)$, to
have a measurable influence in $B$ decays, calculated at the scale
$\mu \simeq m_b$, due to the effects of the renormalization group.
 Qualitatively, this picture also holds in the analysis of
the exclusive non-leptonic $B$ decays discussed by us. We show a typical 
case $B^\pm \to K \pi^\pm$ in Fig.~\ref{C8nsm}, where the branching ratio
for this mode is plotted as a function of the variable
$C_8(m_W)/C_{8}^{SM}(m_W)$. Despite the large range of this
variable, we find that the influence of such a markedly enhanced 
$C_8(m_W)$ on non-leptonic $B$ decays is marginal. In future, we hope that
these matters will be scrutinized much more minutely.
We conclude that the non-leptonic $B$-decays considered here do not require
large enhancements of $C_8(m_W)$, 
or of any other Wilson coefficient, as they are by 
and large compatible with 
data with their SM values.

\section{Summary}

In the first part of this paper we
 have presented a theoretical framework to study two-body
decays of $B$ mesons with two light mesons in the final state. 
First, we took into account the complete NLL corrections
at the partonic level, improving thereby 
previous calculations. 
In particular, we have also 
included the effects of the chromomagnetic penguin operator in 
non-leptonic $B$ decays. These NLL order corrections 
are numerically important in the exclusive decay rates.
Second, to estimate the
hadronic matrix elements we assumed factorization and gave a 
parametrization
for the so-called heavy-to-light transitions. In its most economic 
version, this brings in a single
phenomenological parameter, called $\xi$, which has to be determined by 
comparing the predictions of this model with data.
We have taken a first look at the available CLEO data and estimated that 
our model is compatible within the present theoretical and experimental
errors with data in the range $0\leq \xi \leq 0.5$. With more precise data
one should be able to test our model and see if
within reasonable accuracy  one 
obtains a universal value for this quantity in the heavy-to-light sector.
Alternatively, with more precise data in several decay modes,
we propose to extract the effective coefficients $a_1,...,a_6$ directly
to determine the extent of non-perturbative effects in each one of them.

In the second part of this paper we have applied this framework in the 
analysis of the exclusive two-body
$B$ decays, in which QCD penguins play an important role. Some of these 
decays have been measured recently by the CLEO 
collaboration \cite{Behrens97,Jsmith97,Wuerthwein97}, with which we 
compared
our model calculations; we have also 
predicted the branching ratios for some related decay modes which have not 
been measured yet.
 While the formalism provided here is generally applicable to study
all $B \to PP, ~B\to PV$ and $B \to VV$ decays, we have restricted ourselves
to discussing the four generic cases: $B \to \pi \pi$, $B \to K \pi$, 
$B^\pm \to \omega h^\pm$ and $B^\pm \to (\eta,\eta^\prime)(K^\pm,K^{\ast 
\pm})$.

 In particular, we  have studied   
 at some length the last class, involving the decay $B^\pm \to K^\pm 
\eta^\prime$ and the related ones.
As the $\eta^\prime$ and $\eta$ mesons are composed of 
$u$, $d$ and $s$ quarks, the corresponding decay rates are particularly
sensitive to interference effects among
the several competing amplitudes involving the current-current and the 
QCD-penguins operators, as  
was pointed out by Lipkin some time ago \cite{Lipkin91}. 
In addition, the operators $O_{1,2}^{c}$ which induce transitions of
the form $b \to s (c\bar{c}) \to s(\eta,\eta^\prime)$ have to be included.
Estimates of the latter require a trustworthy evaluation of the
$c\bar{c}$ component in the wave-function of the $\eta^\prime$ and $\eta$
mesons. We have used the mixing formalism involving
the  $(\eta,\eta^\prime,\eta_c)$-complex and data to determine the $c\bar{c}$
contents of these mesons.  
We find that this charm-induced contribution does not dominate the
amplitudes for the processes involving $\eta^\prime$; the decay rate is 
more sensitive to the penguin contributions. 
Our paper provides the complete amplitudes showing
all these individual contributions. This can be used
 in future analyses
of more precise data to determine the $c\bar{c}$ components in $\eta$ and
$\eta^\prime$. 
The estimates presented here with $|f_{\eta^\prime}^{(c)}| = 6$ MeV
yield  ${\cal 
B}(B^\pm \to K^\pm \eta^\prime) = (2 - 4) \times 10^{-5}$. This is
somewhat lower than the central value of the present measurement but 
compatible with the value obtained by fluctuating down the experimental error
by $ 1 \sigma$.  However, a simple answer about the large measured value 
of ${\cal B}(B^\pm \to \eta^\prime K^\pm)$, a question frequently
asked, in terms of a single dominating amplitude is not readily 
available, though the penguins and the singlet component of the
$\eta'$ are certainly at the back of the enhanced branching ratio for
this decay. In our analysis,
we find that the measured rate in the $\eta^\prime K^\pm$ mode is only 
marginally (say a factor 2) larger than our model estimates and given the
theoretical errors there is nothing anomalous about it.
We also expect that the data will evolve with time so as to reduce the 
present discrepancy.

We have made predictions to 
test this interference pattern in the related decays involving $\eta$ 
and $\eta^\prime$. The
resulting decay rates, which also reflect the built-in angular momentum
differences between the states $K^\ast (\eta,\eta^\prime)$ and 
$K (\eta,\eta^\prime)$, show a certain hierarchy among the branching 
ratios. While the other three may turn out
comparable with each other (within a factor 2 -3), we predict:
$$ {\cal B}(B^\pm \to \eta^\prime K^\pm) \gg {\cal B}(B^\pm \to 
\eta^\prime K^{\ast \pm}).$$
The measurement of ${\cal B}(B^\pm \to \eta^\prime K^\pm)$ being the largest
measured so far is in line with our analysis based on the SM. However,
in our SM-based framework it would be
 difficult to accommodate  a much larger branching ratio
${\cal B}(B^\pm \to \eta^\prime K^{\ast \pm})$ vitiating
this hierarchy. 

  The rates for the other decays presented in our analysis are 
also in reasonable
agreement with data, within the presently allowed CKM-parameter space.
Based on our estimates presented here, we expect the decay mode
$B^0(\overline{B^0}) \to \pi^\mp \pi^\pm$ to be measured within a factor 
2 - 3 below the present upper limit.
We point out interesting inferences which present data allows to draw
on the consistency of the SM. In particular, the ratios $R_1$ and $R_2$
involving the $K\pi$ and $\pi \pi$ final states
appear very promising. Present measurements on these ratios are tantalizingly
close to providing independent information on the CKM-Wolfenstein
parameters $\rho$ and $\eta$.
 Their impact on the CKM phenomenology will only
be determined with more precise data, to which we look forward with
animated interest. For the time being, 
the standard model rules OK -- also
in the non-leptonic $B$ decays! 
    
\vspace{2cm}

{\bf Acknowledgements:}

  We are very grateful to Tom Browder, Jim Smith, Tomasz Skwarnicki and 
Frank W\"urthwein for helpful correspondence concerning the CLEO data.
In particular, we thank Jim Smith for sharing his insight on several
points discussed in this report and for asking us incisive questions
related to the CLEO data and our analysis. Helpful discussions with Gustav 
Kramer, Hans K\"uhn, Heiri Leutwyler, Peter Minkowski,
Hubert Simma and Daniel Wyler
on various theoretical issues discussed here are also thankfully acknowledged.
We also thank Nilendra Deshpande, Bhaskar Dutta, Sechul Oh,  Guido 
Martinelli and Amarjit Soni for correspondence and
discussions on the earlier version of this manuscript, and Thorsten
Feldmann and Peter Kroll for sending us an advanced copy of their paper.

Note added in Proofs: Since the publication of this paper, several related
papers have appeared in which some of the issues discussed here are also
investigated \cite{Pakvasa97,DGR97,CT97,DDO97}. 

\begin{figure}[p]
\centerline{
\epsfig{file=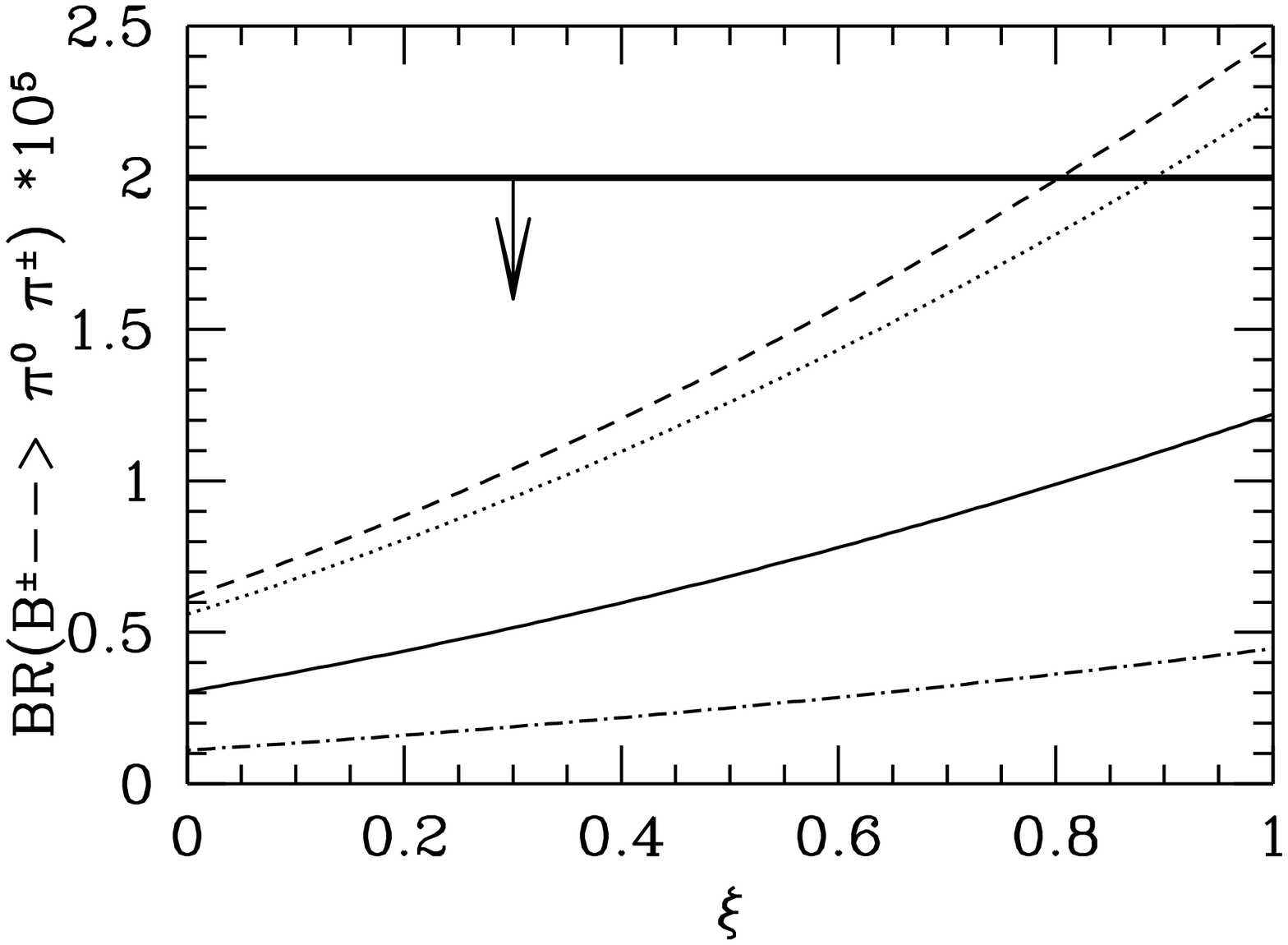,
height=3.0in,angle=0,clip=}
}
\caption[]{
Branching ratio for $B^\pm \to \pi^0 \pi^\pm$  
as a function of $\xi$ for various points in the $(\rho;\eta)$-plane.
Solid curve: $(\rho,\eta)=(0.05,0.36)$; 
dotted curve: $(\rho,\eta)=(-0.20,0.45)$; 
dashed curve: $(\rho,\eta)=(0.30,0.42)$; 
dashed-dotted curve: $(\rho,\eta)=(0.00,0.22)$. 
The horizontal thick solid line (with the arrow) shows the CLEO upper limit
(at 90\% C.L.).
\label{Bpmpipi}}
\end{figure}

\begin{figure}[p]
\centerline{
\epsfig{file=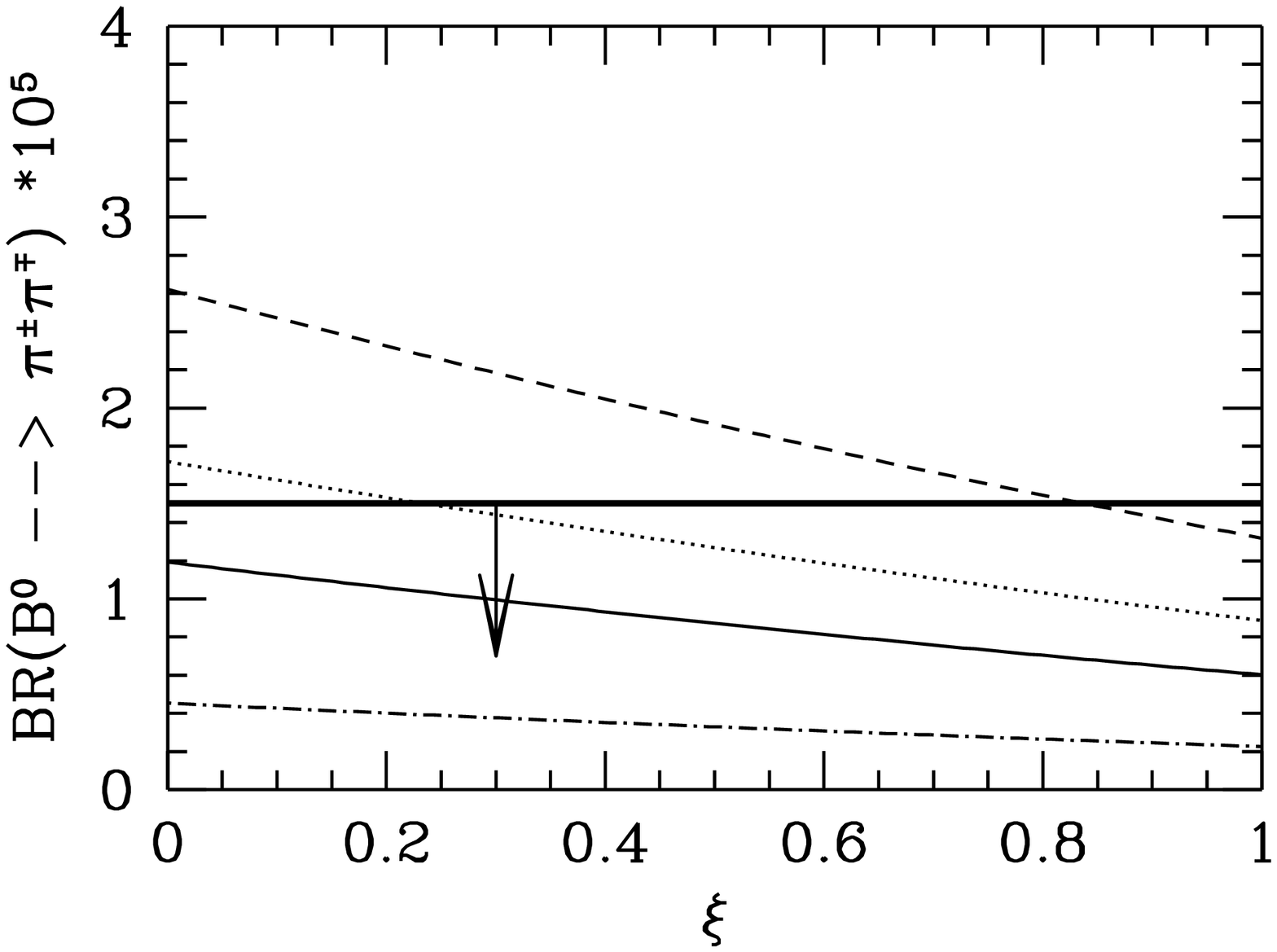,
height=3.0in,angle=0,clip=}
}
\caption[]{
As Fig.~\ref{Bpmpipi} but for the process 
$B^0 (\overline{B^0}) \to \pi^\pm \pi^\mp$.
\label{Bzpippim}}
\end{figure}

\begin{figure}[p]
\centerline{
\epsfig{file=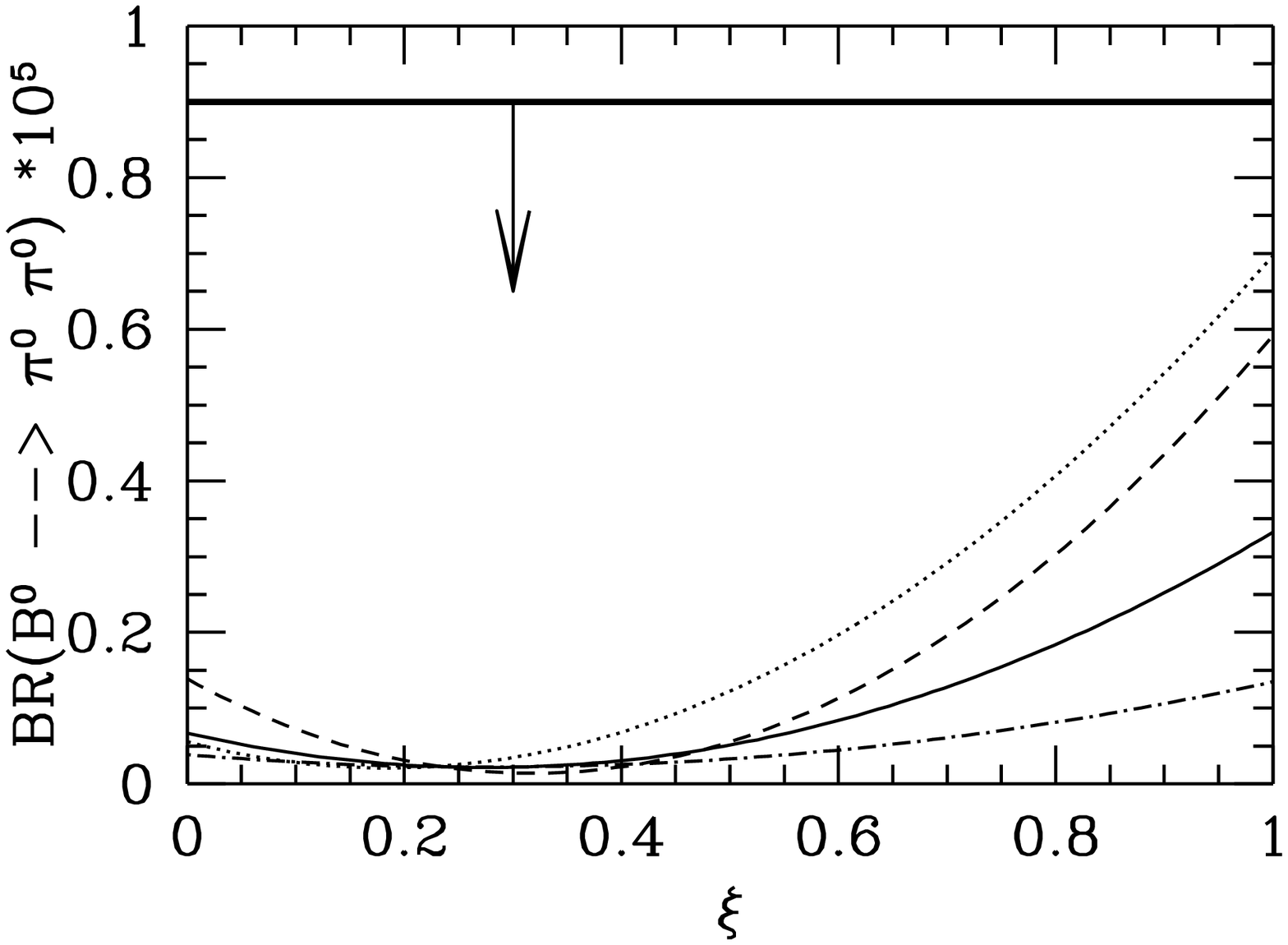,
height=3.0in,angle=0,clip=}
}
\caption[]{
As Fig.~\ref{Bpmpipi} but for the process 
$B^0 (\overline{B^0}) \to \pi^0 \pi^0$.
\label{Bzpi0pi0}}
\end{figure}

\begin{figure}[p]
\centerline{
\epsfig{file=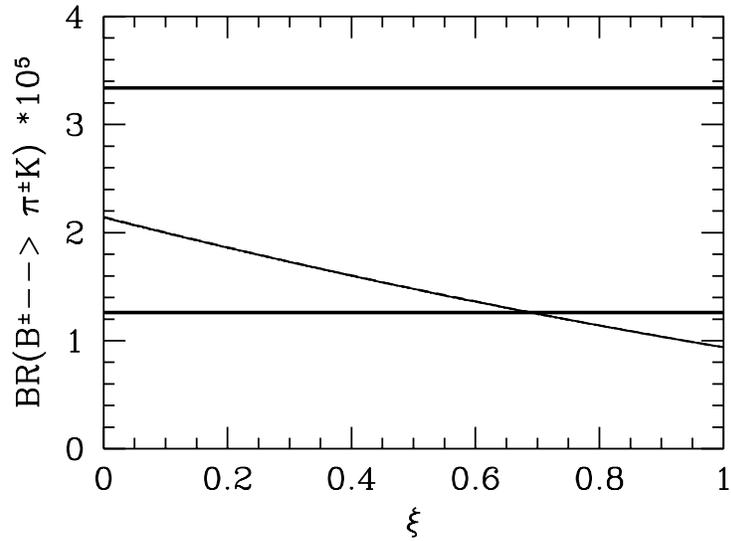,
height=3.0in,angle=0,clip=}
}
\caption[]{
Branching ratio for $B^\pm \to \pi^\pm K  $ 
as a function of $\xi$ for various points in the $(\rho,\eta)$-plane.
Solid line: $(\rho,\eta)=(0.05,0.36)$; 
dotted line: $(\rho,\eta)=(-0.20,0.45)$; 
dashed line: $(\rho,\eta)=(0.30,0.42)$; 
dash-dotted line: $(\rho,\eta)=(0.00,0.22)$. The thick 
solid lines show the CLEO measurement (with $\pm 1\sigma$ errors). 
\label{Bpmpik}}
\end{figure}

\begin{figure}[p]
\vspace{0.10in}
\centerline{
\epsfig{file=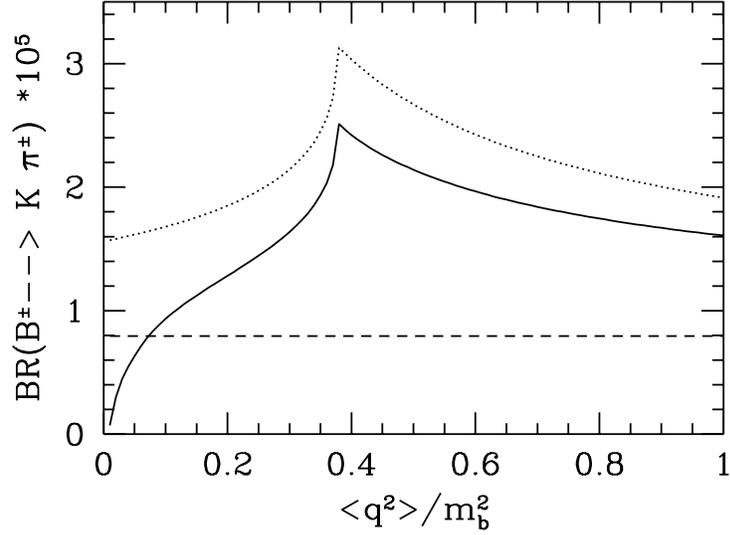,
height=3.0in,angle=0,clip=}
}
\vspace{0.08in}
\caption[]{${\cal B}(B^\pm \to \pi^\pm K)$
as a function of $\bra q^2\ket/m_b^2$. The dashed line
corresponds to the LL approximation. The solid line and the 
dotted line correspond both 
to the Wilson coefficients evaluated in the NLL approximation; the solid 
line 
takes into account the penguin diagrams of all the four-Fermi operators
and the tree level matrix element of $O_8$, while the dotted line
takes into account the penguin diagrams associated with 
the four-Fermi operators $O^c_{1,2}$ only.
\label{Bkpinll}}
\end{figure}

\begin{figure}[p]
\centerline{
\epsfig{file=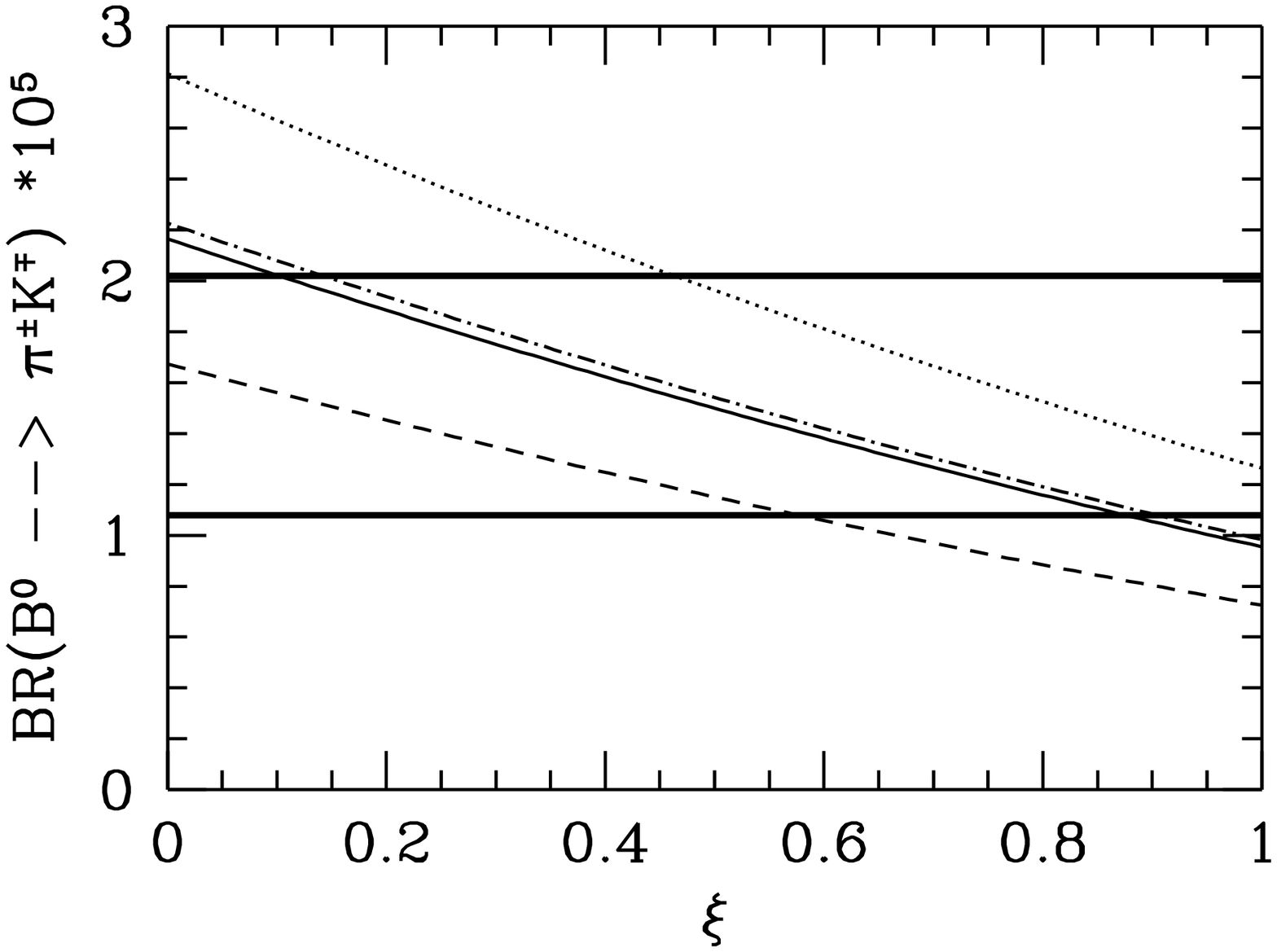,
height=3.0in,angle=0,clip=}
}
\caption[]{
As Fig.~\ref{Bpmpik} but for the process 
$B^0 (\overline{B^0}) \to \pi^\pm K^\mp$.
\label{Bzpik}}
\end{figure}

\begin{figure}[p]
\vspace{0.10in}
\centerline{
\epsfig{file=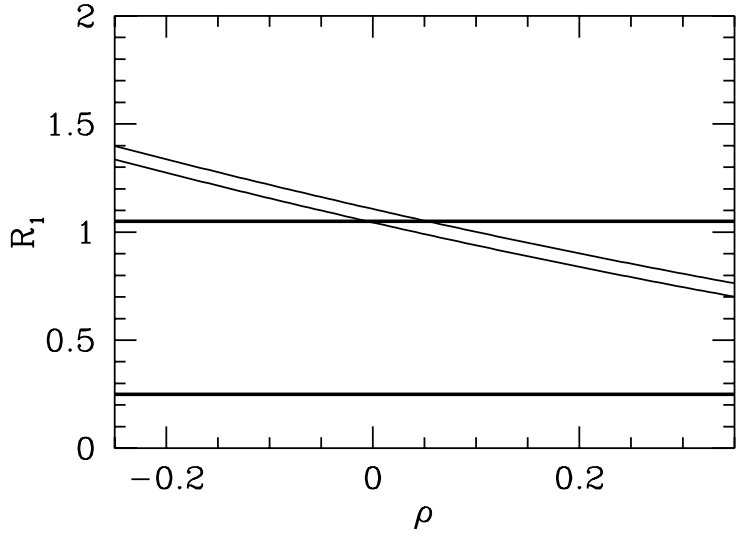,
height=3.0in,angle=0,clip=}
}
\vspace{0.08in}
\caption[]{
The ratio 
$R_1={\cal B}(B^0 (\overline{B^0}) 
\to \pi^\pm K^\mp)/{\cal B}(B^\pm \to K \pi^\pm)$
as a function of the Wolfenstein parameter $\rho$, for $\eta=0.25$ (lower
curve) and $\eta=0.52$ (upper curve). The curves are drawn for $\xi=0$. 
The horizontal thick solid lines
show the CLEO measurement (with $\pm 1\sigma$ errors).  
\label{R1}}
\end{figure}

\begin{figure}[p]
\centerline{
\epsfig{file=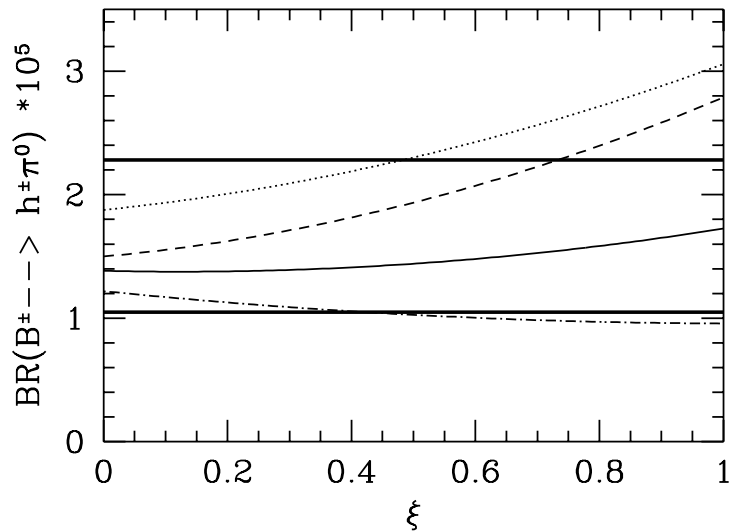,
height=3.0in,angle=0,clip=}
}
\caption[]{
As Fig.~\ref{Bpmpik} but for the process $B^\pm \to h^\pm \pi^0$
($h=K,\pi$).
\label{Bpmpihpm}}
\end{figure}

\begin{figure}[p]
\centerline{
\epsfig{file=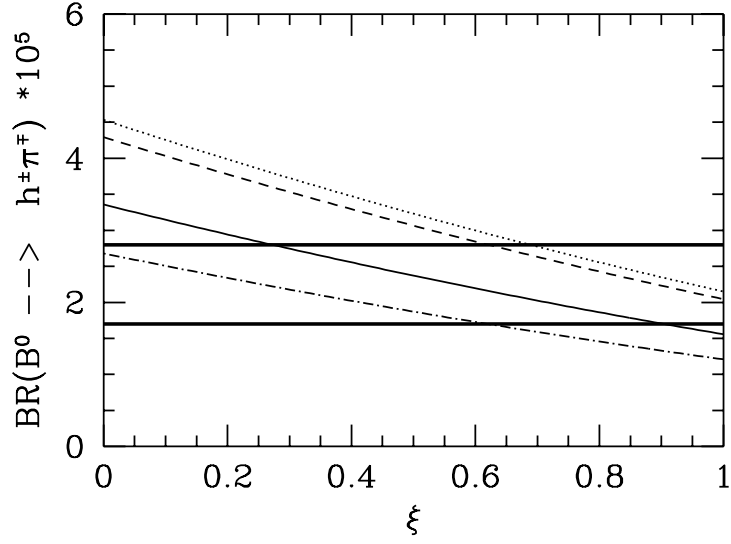,
height=3.0in,angle=0,clip=}
}
\caption[]{
As Fig.~\ref{Bpmpik} but for 
$B^0 (\overline{B^0}) \to h^\pm \pi^\mp  $ .
\label{Bzhpmpi}}
\end{figure}

\begin{figure}[p]
\centerline{
\epsfig{file=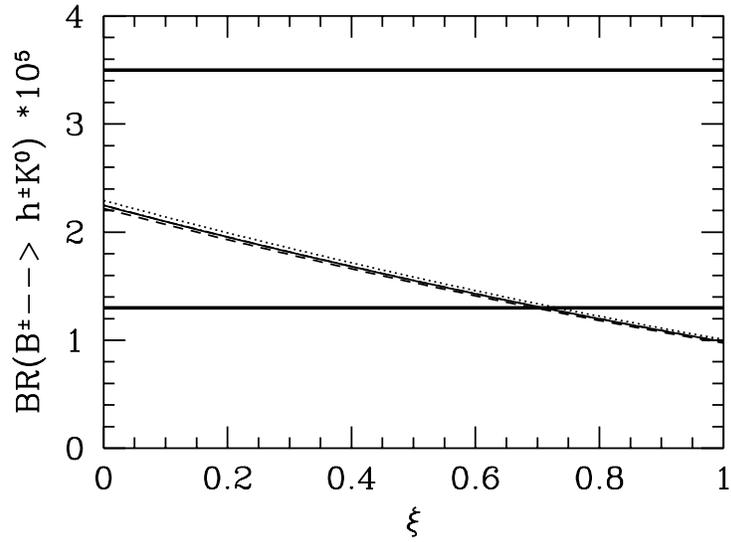,
height=3.0in,angle=0,clip=}
}
\caption[]{
As Fig.~\ref{Bpmpik} but for the process $B^\pm \to h^\pm K$.
\label{Bpmhpmk}}
\end{figure}

\begin{figure}[p]
\vspace{0.10in}
\centerline{
\epsfig{file=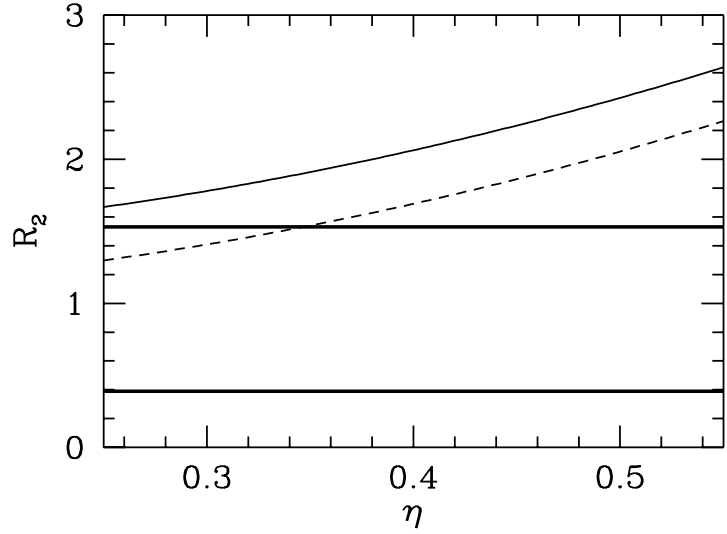,
height=3.0in,angle=0,clip=}
}
\vspace{0.08in}
\caption[]{
The ratio 
$R_2={\cal B}(B^0 (\overline{B^0}) 
\to h^\pm \pi^\mp)/{\cal B}(B^\pm \to K \pi^\pm)$
as a function of the Wolfenstein parameter $\eta$, for $\rho=0.05$ 
(dashed curve)
and $\rho=0.35$ (solid curve). For $\rho=-0.25$ the corresponding curve
is almost identical to the curve for $\rho=0.35$
The curves are drawn for $\xi=0$.  
The horizontal thick solid lines
show the CLEO measurement (with $\pm 1\sigma$ errors).  
\label{R2}}
\end{figure}

\begin{figure}[p]
\vspace{0.10in}
\centerline{
\epsfig{file=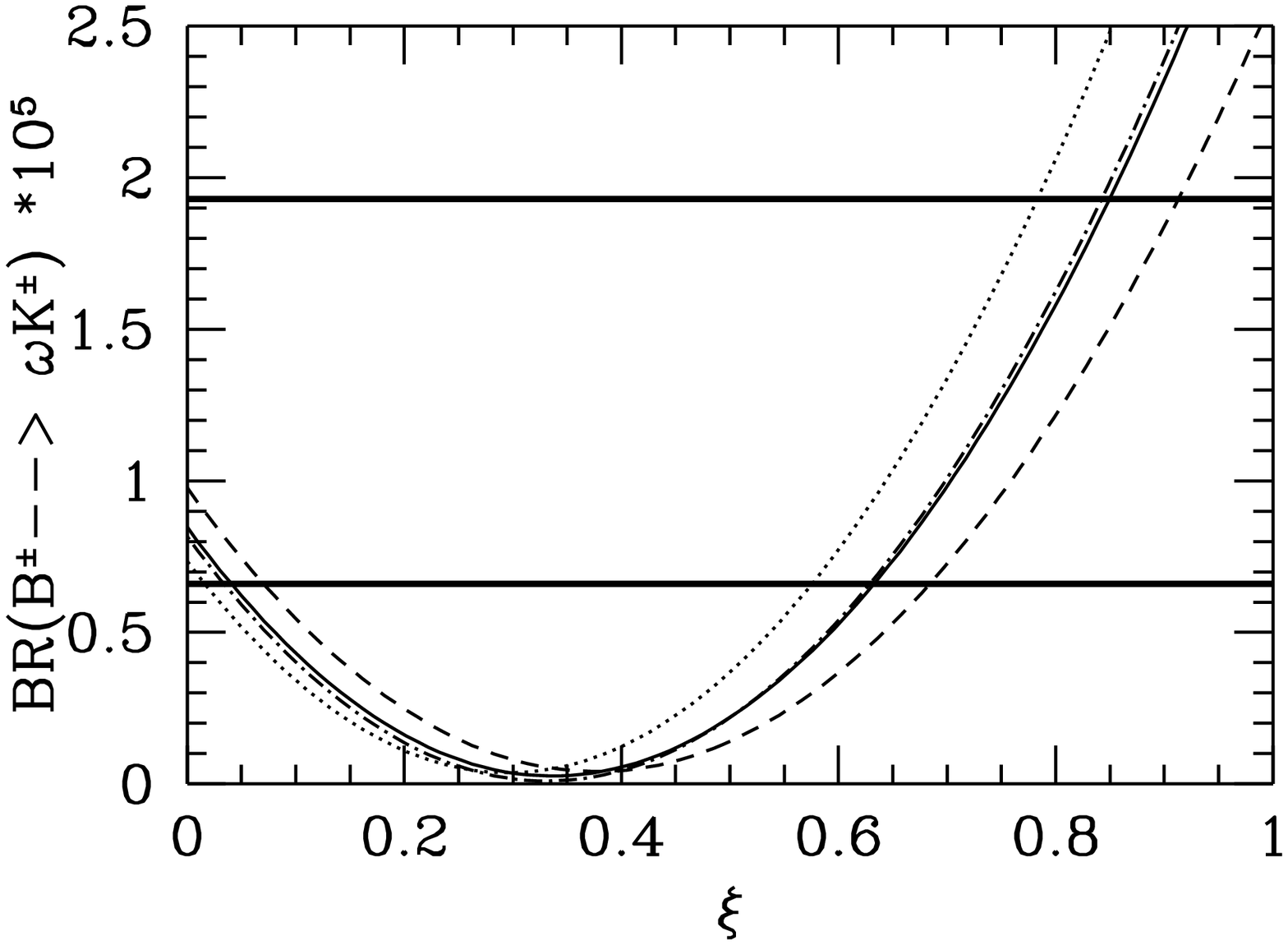,
height=3.0in,angle=0,clip=}
}
\vspace{0.08in}
\caption[]{
As Fig.~\ref{Bpmpik} but for the process $B^\pm \to \omega K^\pm$.
\label{Bpmomegakpm}}
\end{figure}

\begin{figure}[p]
\vspace{0.10in}
\centerline{
\epsfig{file=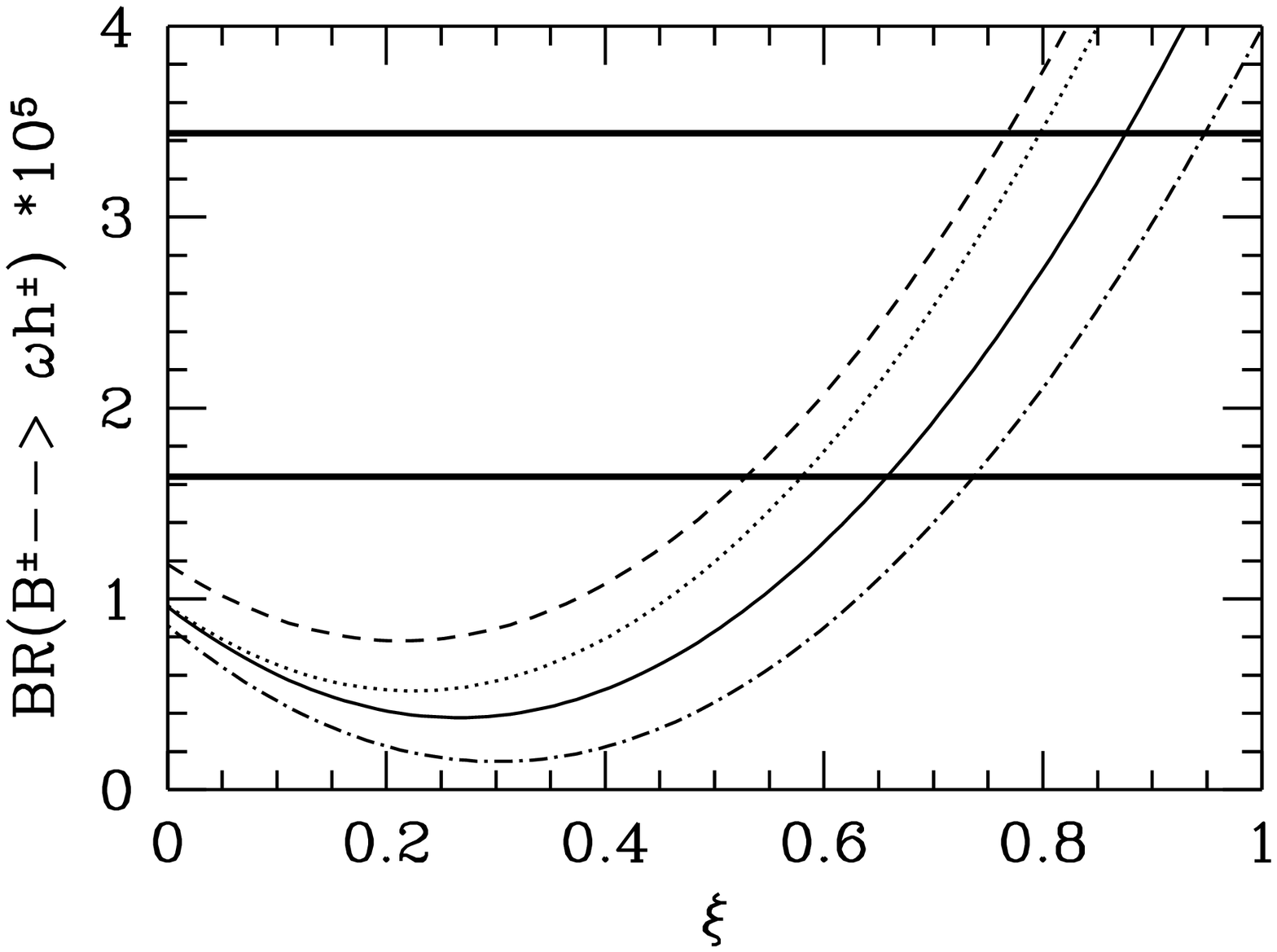,
height=3.0in,angle=0,clip=}
}
\vspace{0.08in}
\caption[]{
As Fig.~\ref{Bpmpik} but for the process $B^\pm \to \omega h^\pm.$
\label{Bpmomegahpm}}
\end{figure}

\begin{figure}[p]
\vspace{0.10in}
\centerline{
\epsfig{file=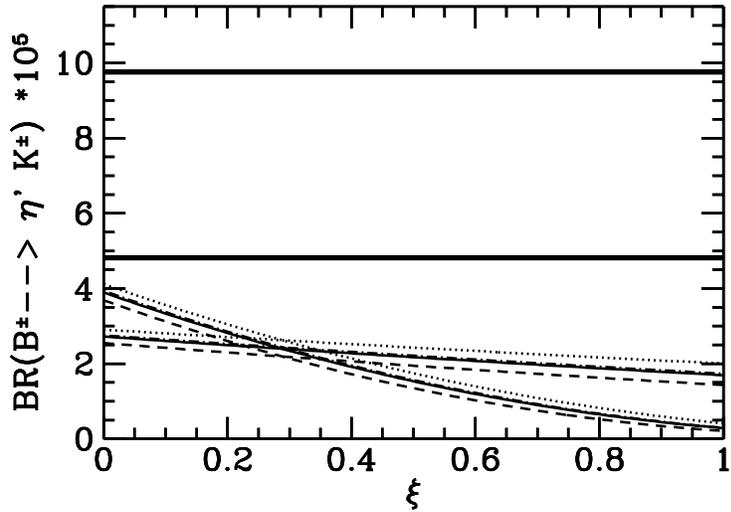,
height=3.0in,angle=0,clip=}
}
\vspace{0.08in}
\caption[]{
Branching ratio for $B^\pm \to \eta' K^\pm $
as a function of $\xi$ for the same points in the $(\rho,\eta)$-plane
as in Fig.~\ref{Bpmpik}. The upper (lower)  set of curves close to $\xi=0$ 
corresponds to the positive (negative) solution for $f_{\eta'}^{(c)}$. 
The horizontal thick 
solid lines show the CLEO measurement (with $\pm 1\sigma$  errors). 
\label{Bpmetapkpm}}
\end{figure}

\clearpage

\begin{figure}[p]
\vspace{0.10in}
\centerline{
\epsfig{file=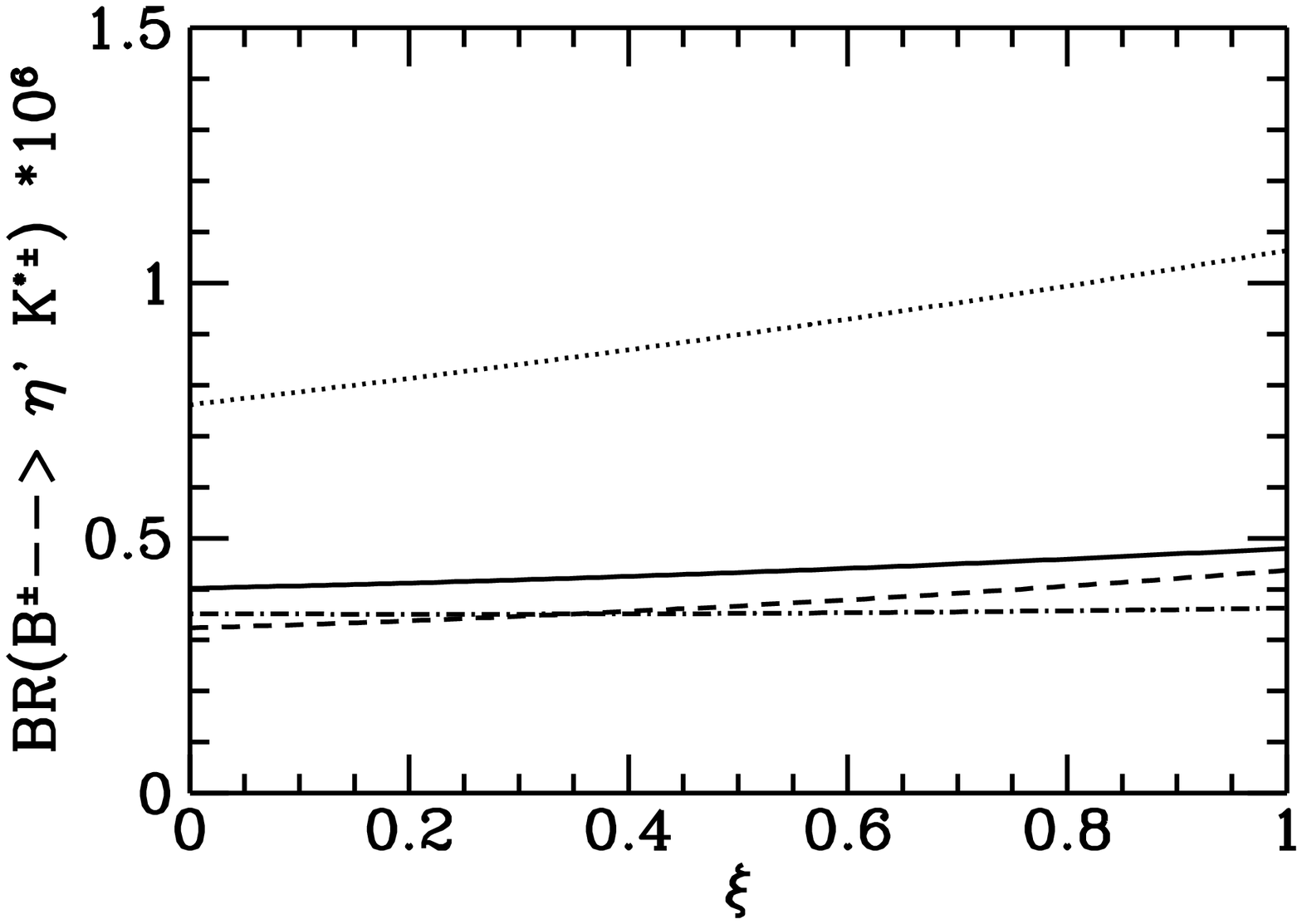,
height=3.0in,angle=0,clip=}
}
\vspace{0.08in}
\caption[]{
Branching ratio for $B^\pm \to \eta' K^{\pm *} $
as a function of $\xi$ for the same points in the $(\rho,\eta)$-plane
as in Fig.~\ref{Bpmpik}. All curves 
correspond to the value $f_{\eta'}^{(c)}=-5.8$ MeV. 
The upper limit from CLEO is $2.9\times 10^{-4}$ at 90\% C.L.. 
\label{Bpmetapkstpma}}
\end{figure}

\begin{figure}[p]
\vspace{0.10in}
\centerline{
\epsfig{file=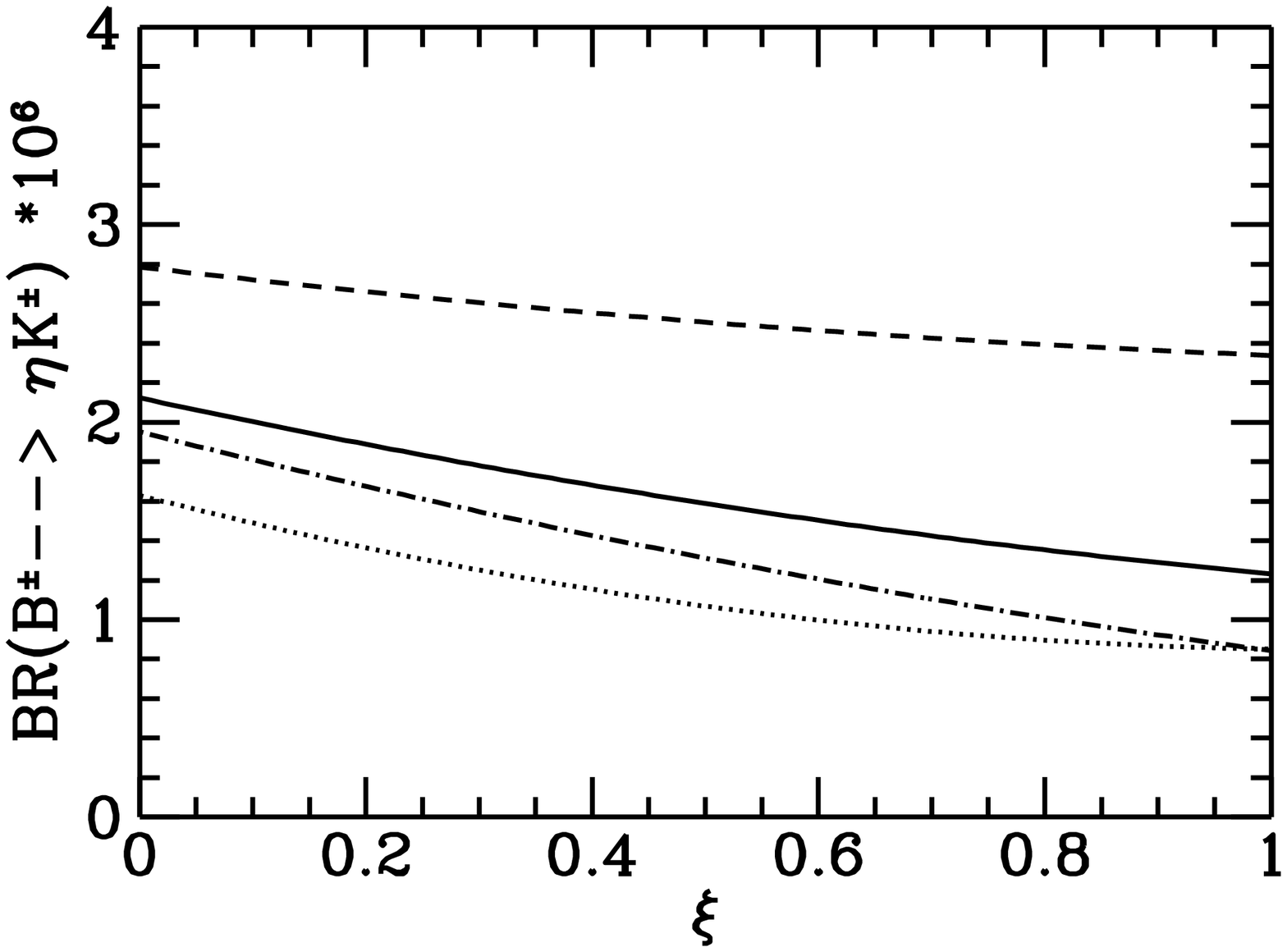,
height=3.0in,angle=0,clip=}
}
\vspace{0.08in}
\caption[]{
Branching ratio for $B^\pm \to \eta K^{\pm } $
as a function of $\xi$ for the same points in the $(\rho,\eta)$-plane
as in Fig.~\ref{Bpmpik}. All curves 
correspond to the value $f_{\eta}^{(c)}=-0.93$ MeV. 
The upper limit from CLEO is $8 \times 10^{-6}$ at 90\% C.L.. 
\label{Bpmetakpm}}
\end{figure}

\begin{figure}[p]
\vspace{0.10in}
\centerline{
\epsfig{file=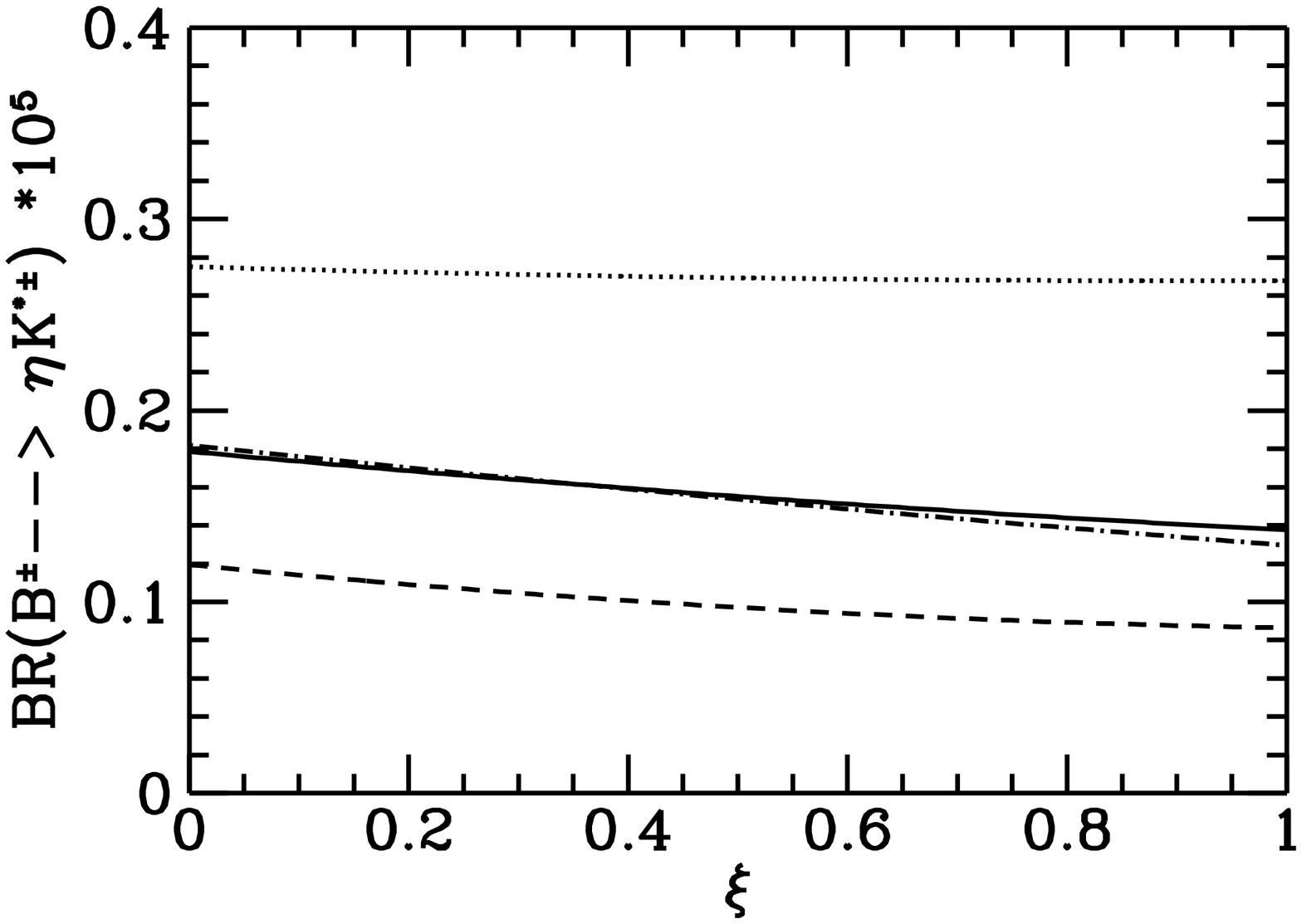,
height=3.0in,angle=0,clip=}
}
\vspace{0.08in}
\caption[]{
Branching ratio for $B^\pm \to \eta K^{\pm *} $
as a function of $\xi$ for the same points in the $(\rho,\eta)$-plane
as in Fig.~\ref{Bpmpik}. All curves 
correspond to the value $f_{\eta}^{(c)}=-0.93$ MeV. 
The upper limit from CLEO is $2.4 \times 10^{-4}$ at 90\% C.L.. 
\label{Bpmetakstpm}}
\end{figure}

\begin{figure}[p]
\vspace{0.10in}
\centerline{
\epsfig{file=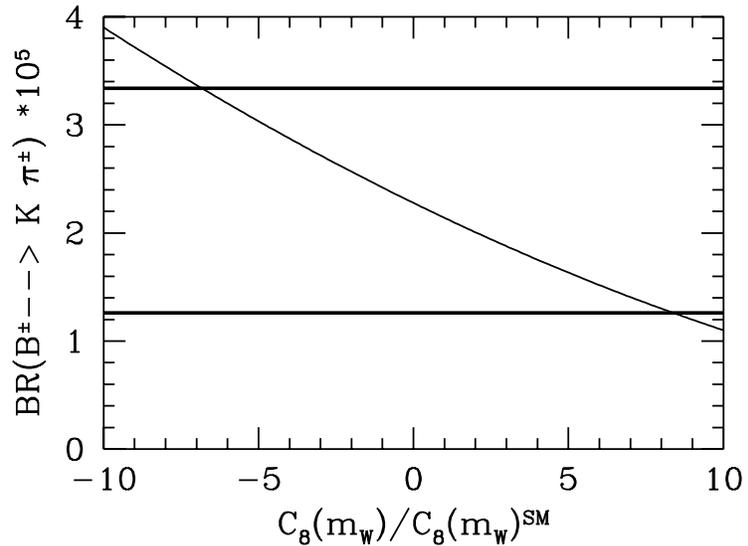,
height=3.0in,angle=0,clip=}
}
\vspace{0.08in}
\caption[]{${\cal B}(B^\pm \to K \pi^\pm)$ as a function of 
$C_8(m_W)/C^{SM}_8(m_W)$, where $C_8(m_W)$ is the Wilson coefficient 
of $O_8$ including
new physics, while $C^{SM}_8(m_W)$ is the standard model value.
We use $\xi=0$ and $(\rho,\eta)=(0.05,0.36)$. 
\label{C8nsm}}
\end{figure}

\end{document}